\documentclass{aa}  
\usepackage{natbib}
\bibpunct{(}{)}{;}{a}{}{,}
\usepackage[colorlinks=true,linkcolor=blue,citecolor=blue,urlcolor=blue]{hyperref} 
\usepackage{graphicx}
\usepackage{txfonts}
\usepackage{orcidlink}
\usepackage{placeins}
\usepackage{float}

\begin{document} 
\authorrunning{Facchini, P., et al.}
\title{Isolated massive star candidates in NGC 4242 with GULP}

\author{
Pietro~Facchini \orcidlink{0009-0003-6044-3989} \inst{\ref{ari}} \and
Eva~K.~Grebel \orcidlink{0000-0002-1891-3794} \inst{\ref{ari}} \and
Anna~Pasquali \orcidlink{0000-0001-5171-5629} \inst{\ref{ari}} \and
Elena~Sabbi \orcidlink{0000-0003-2954-7643} \inst{\ref{gemini},\ref{stsci}} \and
Beena~Meena \orcidlink{0000-0001-8658-2723} \inst{\ref{stsci}} \and
Varun~Bajaj \inst{\ref{stsci}} \and
John~S.~Gallagher~III \orcidlink{0000-0001-8608-0408} \inst{\ref{wisconsin}} \and
Bruce~G.~Elmegreen \orcidlink{0000-0002-1723-6330} \inst{\ref{katonah}}\and
Luciana~Bianchi \orcidlink{0000-0001-7746-5461} \inst{\ref{jhu}} \and
Angela~Adamo \orcidlink{0000-0002-8192-8091} \inst{\ref{stockholm}} \and
Daniela~Calzetti \orcidlink{0000-0002-5189-8004} \inst{\ref{mass}} \and
Michele~Cignoni \orcidlink{0000-0001-6291-6813} \inst{\ref{unipisa},\ref{infnpisa},\ref{inafbologna}} \and
Paul~A.~Crowther \orcidlink{0000-0001-6000-6920} \inst{\ref{sheffield}} \and
Jan~J.~Eldridge \orcidlink{0000-0002-1722-6343} \inst{\ref{aukland}} \and
Mario~Gennaro \orcidlink{0000-0002-5581-2896} \inst{\ref{stsci}} \and
Ralf~S.~Klessen \orcidlink{0000-0002-0560-3172} \inst{\ref{ita},\ref{unihd},\ref{harvard},\ref{radcliffe}} \and
Linda~J.~Smith \orcidlink{0000-0002-0806-168X} \inst{\ref{stsci}} \and
Aida~Wofford \orcidlink{0000-0001-8289-3428} \inst{\ref{unam}} \and
Peter~Zeidler \orcidlink{0000-0002-6091-7924} \inst{\ref{aura}}
}

\institute{
Astronomisches Rechen-Institut, Zentrum f\"ur Astronomie der Universit\"at Heidelberg, M\"onchhofstr. 12-14, D-69120 Heidelberg, Germany \label{ari}\\
email:\href{mailto:pietro.facchini@stud.uni-heidelberg.de}{pietro.facchini@stud.uni-heidelberg.de}
\and
Gemini Observatory/NSFs NOIRLab, 950 N. Cherry Ave., Tucson, AZ 85719, USA \label{gemini}
\and
Space Telescope Science Institute, 3700 San Martin Dr, Baltimore, MD 21218, USA \label{stsci}
\and
Department of Astronomy, University of Wisconsin-Madison, 475 North Charter St. Madison, WI 53706 USA \label{wisconsin}
\and
Katonah, NY 10536, USA \label{katonah}
\and
Dept. of Physics and Astronomy, the Johns Hopkins University \label{jhu}
\and
Department of Astronomy, Oskar Klein Centre, Stockholm University, AlbaNova University Center, SE-106 91 Stockholm, Sweden \label{stockholm}
\and
Department of Astronomy, University of Massachusetts Amherst, 710 North Pleasant Street, Amherst, MA 01003, USA \label{mass}
\and
Dipartimento di Fisica, Università di Pisa, Largo Bruno Pontecorvo 3, 56127, Pisa, Italy \label{unipisa}
\and
INFN, Largo B. Pontecorvo 3, 56127, Pisa, Italy \label{infnpisa}
\and
INAF - Osservatorio di Astrofisica e Scienza dello Spazio di Bologna, Via Piero Gobetti 93/3, 40129, Bologna, Italy \label{inafbologna}
\and
Department of Physics \& Astronomy, University of Sheffield, Hounsfield Road, Sheffield, S3 7RH, United Kingdom \label{sheffield}
\and
Department of Physics, The University of Auckland, Private Bag 92019, Auckland, New Zealand \label{aukland}
\and
Universit\"{a}t Heidelberg, Zentrum f\"{u}r Astronomie, Institut f\"{u}r Theoretische Astrophysik, Albert-Ueberle-Str.\ 2, 69120 Heidelberg, Germany \label{ita}
\and
Universit\"{a}t Heidelberg, Interdisziplin\"{a}res Zentrum f\"{u}r Wissenschaftliches Rechnen, Im Neuenheimer Feld 225, 69120 Heidelberg, Germany \label{unihd}
\and
Harvard-Smithsonian Center for Astrophysics, 60 Garden Street, Cambridge, MA 02138, U.S.A. \label{harvard}
\and
Elizabeth S. and Richard M. Cashin Fellow at the Radcliffe Institute for Advanced Studies at Harvard University, 10 Garden Street, Cambridge, MA 02138, U.S.A. \label{radcliffe}
\and
Instituto de Astronom\'{i}a, Universidad Nacional Aut\'{o}noma de M\'{e}xico, Unidad Acad\'{e}mica en Ensenada, Km 103 Carr. Tijuana-Ensenada, Ensenada 22860, M\'{e}xico \label{unam}
\and
AURA for the European Space Agency (ESA), ESA Office, Space Telescope Science Institute, 3700 San Martin Drive, Baltimore, MD 21218, USA \label{aura}
}

\date{Received XXX; accepted YYY}

  \abstract
   {There is considerable debate on how massive stars form, including whether a high-mass star must always form with a population of low-mass stars or whether it can also form in isolation. Massive stars found in the field are often considered to be runaways from star clusters or OB associations. However, there is evidence in the Milky Way and the Small Magellanic Cloud of high-mass stars that appear isolated in the field and cannot be related to any known star cluster or OB association. Studies of more distant galaxies have been lacking so far. }
   {In this work, we identified massive star candidates that appear isolated in the field of the nearby spiral galaxy NGC 4242 (distance: 5.3 Mpc), to explore how many candidates for isolated star formation we find in a galaxy outside the Local Group.}
   {We identified 234 massive ($M_{ini}\geq15M_{\odot}$) and young ($\leq 10$~Myr) field stars in NGC $4242$ using the Hubble Space Telescope's Solar Blind Channel of the Advanced Camera for Surveys, the UVIS channel of the Wide Field Camera~3 from the Galaxy UV Legacy Project (GULP) and optical data from the Legacy ExtraGalactic UV Survey (LEGUS). We investigated the surroundings of our targets within the range of projected distances expected for runaway stars, $74$~pc and $204$~pc.}
   {We find that between $9.8\%$ and $34.6\%$ of our targets have no young stellar groups or massive stars within the threshold radii, making them appear isolated. This fraction reduces to $3.2\%-11.5\%$ when we consider the total number of massive stars expected from the observed UV star formation rate.}
   {Our results show that there is a small population of young and massive, potentially isolated field stars in NGC $4242$.}

   \maketitle

\section{Introduction}\label{sec:introduction}
Although small in number, massive stars are powerful engines that affect the local star formation activity of the host galaxy, quenching star formation in their neighborhood (due to stellar winds or supernova explosions, e.g. \citealt{Westmoquette2008}) or triggering new star formation episodes at larger distances (a supernova explosion can compress the interstellar medium, triggering star formation, e.g. \citealt{Deharveng2009}). Moreover, they are key agents of galactic chemical evolution.

\citet{Lada&Lada2003} show that stars usually form in clustered structures via the collapse and fragmentation of a giant molecular cloud. In massive rich OB clusters, \citet{Garcia2001} further show that O stars are frequently found in multiple or binary systems. Given their short lifetimes and their usual low velocity, high-mass stars usually do not have enough time to move away from their parental star clusters or OB associations, where they formed. However, \citet{Oey2004} show that up to $35\%$ of the massive stars in the Small Magellanic Cloud (SMC) are located in the field, while \citet{Gies1987} finds a fraction of $22\%$ for the Milky Way (MW). Do these findings imply that massive stars can also form in isolation?

The competitive accretion model of star formation predicts that the formation of a high-mass star requires the presence of a population of low-mass stars \citep{Bonnell2004}. In this scenario, the fragmentation of the parental molecular cloud produces only low-mass cores, some of which compete to accrete the remaining gas, producing high-mass stars. The mass of the most massive object ($m_{max}$) is then connected to the mass of the resulting cluster ($M_{cl}$) via $M_{cl} \propto m_{max}^{1.5}$ \citep{Bonnell2004}. In contrast, the core accretion model of star formation allows high-mass stars to form in isolation \citep{Krumholz2009}. In this case, the heat produced by the accretion of gas onto a stellar core can be absorbed by the cloud, and this can prevent the further fragmentation of the parental molecular gas cloud. If the column density of the latter is high enough, a monolithic collapse of the clump could happen, leading to the formation of a single massive star. The feedback produced by massive stars has been proposed by \citet{Elmegreen1977} to trigger the formation of massive stars. In this model, the ionization front coming from an already formed group of massive stars compresses the neutral gas layers, heating them and inducing the formation of a new group of stars \citep[e.g.][]{Zavagno2006}. The higher thermal Jeans mass in the irradiated layers would favor the formation of more massive stars, whereas the formation of low-mass stars will proceed independently and spread out over the cloud. However, this proposal contradicts observations in the Scorpius-Centaurus OB association \citep[Sco-Cen,][]{Preibisch&Zinnecker2007} which show that low-mass T Tauri stars are coeval with the OB population. More recently, \citet{Ratzenbock2023} reconstructed the star formation history of Sco-Cen using Gaia data, revealing a coherent and sequential pattern of clusters in space and time reminiscent of the scenario proposed by \citet{Elmegreen1977}. However, they did not address the initial mass function of the sequentially formed clusters nor whether low- and high-mass stars could originate from distinct formation channels.
In the context of ionization-triggered star formation as in \citet{Elmegreen1977}, it is plausible that an early-formed single massive star suppresses subsequent star formation through its feedback, thus ending up isolated.
We note that the model of \citet{Elmegreen1977} investigates only the formation of the new group of massive stars, leaving the formation process of the first massive stars undisputed. A more detailed review of high-mass formation theories and observations is presented in \citet{Zinnecker2007}.

Previous investigations of the OB field star population in the SMC have shown that the field population is dominated by runaway OB stars from parental star clusters or OB associations \citep{DorigoJones2020}. In fact, OB stars can be ejected from their parental star cluster/OB association via two main possible mechanisms: the dynamical ejection scenario (DES), in which strong gravitational interactions in the cluster core can slingshot a massive star into the field \citep{Blaauw1961,Poveda1967}, and the binary supernova scenario (BSS), in which the supernova explosion of the most massive star in a binary system can eject its high-mass companion \citep{Blaauw1961,Renzo2019}. More recently, \citet{Polak2024} have proposed a third scenario, the subcluster ejection scenario (SCES). In the hierarchical formation of a star cluster, subclusters merge and form a single cluster. However, during the merging process, a subcluster can fall into the contracting central gravitational potential of the proto cluster, being disrupted by the gravitational force, and ejecting its own stars as runaways in tidal tails.

\begin{table}
\caption{Log of observations}
\label{tab:Table filters}
\centering
\begin{tabular}{cccc}
\hline\hline
Obs. Date & Instrument & Filter & Total Exposure Time \\
(yy mm dd) & & & (s) \\
\hline
$13$ \text{Feb} $2021$ & HST/ACS & F$150$LP & $5400$ \\
$29$ \text{Apr} $2021$ & HST/WFC$3$ & F$218$W & $2160$ \\
$12$ \text{Apr} $2014$ & HST/WFC$3$ & F$275$W & $2376$ \\
$12$ \text{Apr} $2014$ & HST/WFC$3$ & F$336$W & $1116$ \\
$12$ \text{Apr} $2014$ & HST/WFC$3$ & F$438$W & $962$ \\
$12$ \text{Apr} $2014$ & HST/WFC$3$ & F$555$W & $1140$\\
$12$ \text{Apr} $2014$ & HST/WFC$3$ & F$814$W & $986$\\
\hline
\end{tabular}
\tablefoot{Observation dates, instruments, filters and total exposure times for the used drizzled images.}
\end{table}

By definition, the minimum $3$D peculiar velocity of a runaway is $30~\text{km}~\text{s}^{-1}$ \citep{Gies1987,Hoogerwerf2001}, which translates into a transverse $2$D velocity of $24~\text{km}~\text{s}^{-1}$ assuming equal velocity components in space \citep{DorigoJones2020}. These velocities imply that the runaway star can travel from tens to hundreds of parsec from its birthplace. Thus, understanding how many O-type stars in the field could be interpreted as runaways and traced back to their parental cluster as opposed to the number of truly isolated massive stars that formed in-situ in isolation becomes crucial in star formation theory and evolution.

Observations in the Milky Way suggest that $4\%\pm2\%$ of all O stars may be truly isolated from known star clusters/OB associations and can not be attributed to runaways \citep{deWit2004,deWit2005}.
Recently, \citet{Zarricueta2023} observed an O3If* that appears isolated in the field, with no evidence of clustering of low-mass stars. Moreover, they excluded the runaway nature from its 2D kinematics.
In the Small Magellanic Cloud, \citet{Oey2013} analyzed $14$ isolated OB stars, finding that $9$ stars show a small overdensity of low-mass stars around them, while the remaining $5$ have no clustering and are candidates for in-situ star formation. \citet{Lamb2010}, analyzing $8$ possible isolated OB stars in the SMC, found that $3$ of them show clustering of low-mass stars, $2$ were runaways and $3$ remained isolated. \citet{Vargas-Salazar2020} found that $4-5\%$ of their field O-type star sample in the SMC were not runaways but were surrounded by small clustering of low-mass stars. Thus, they argued that the O stars that appear isolated are, in fact, the "tip of the iceberg" of very sparse star clusters.

In this work, we use far-ultraviolet to optical photometry to identify possible massive and young field stars in NGC $4242$, and analyze their surroundings, studying the isolation of each target with respect to young star clusters, OB associations or other massive stars. The paper is organized as follows. Section~\ref{sec:NGC4242} describes the observations and catalogs; the color-magnitude diagrams and extinction law analysis are presented in Section~\ref{sec:Massive stars section}, while the analysis on the isolation is presented in Section~\ref{sec:Isolation analysis}. The discussion and the conclusions are presented in Section~\ref{sec:Discussion} and \ref{sec:Conclusions}.

\section{NGC 4242 and data description}\label{sec:NGC4242}

We study the unbarred spiral galaxy (SAd) NGC $4242$ located at $\text{RA(J2000)}=12\text{:}17\text{:}30.18$, $\text{DEC(J2000)}=+45\text{:}37\text{:}9.51$. The galaxy has a photometric estimated metallicity of $Z=0.02$, and is located at a distance of $5.3$ Mpc \citep{Sabbi2018}.

The galaxy is part of the star forming galaxies studied by \citet{Lee2011}, who derived a star formation rate (SFR) over the past $100$~Myr of $\sim0.09 M_{\odot}~\text{yr}^{-1}$ using the far-ultraviolet emission detected by GALEX, corrected for the Galactic dust attenuation using the maps of \citet{Schlegel1998}, and a distance of $7.43$ Mpc \citep{Kennicutt2008}.

The face-on orientation of NGC $4242$ makes the galaxy an excellent benchmark for the search for isolated massive stars. A more strongly inclined orientation, instead, would have required probing the distribution of massive stars through the stellar disc, likely rich in dust, which would have considerably affected the UV observations used in this paper.

\begin{figure}
   \centering
   \resizebox{\hsize}{!}{\includegraphics{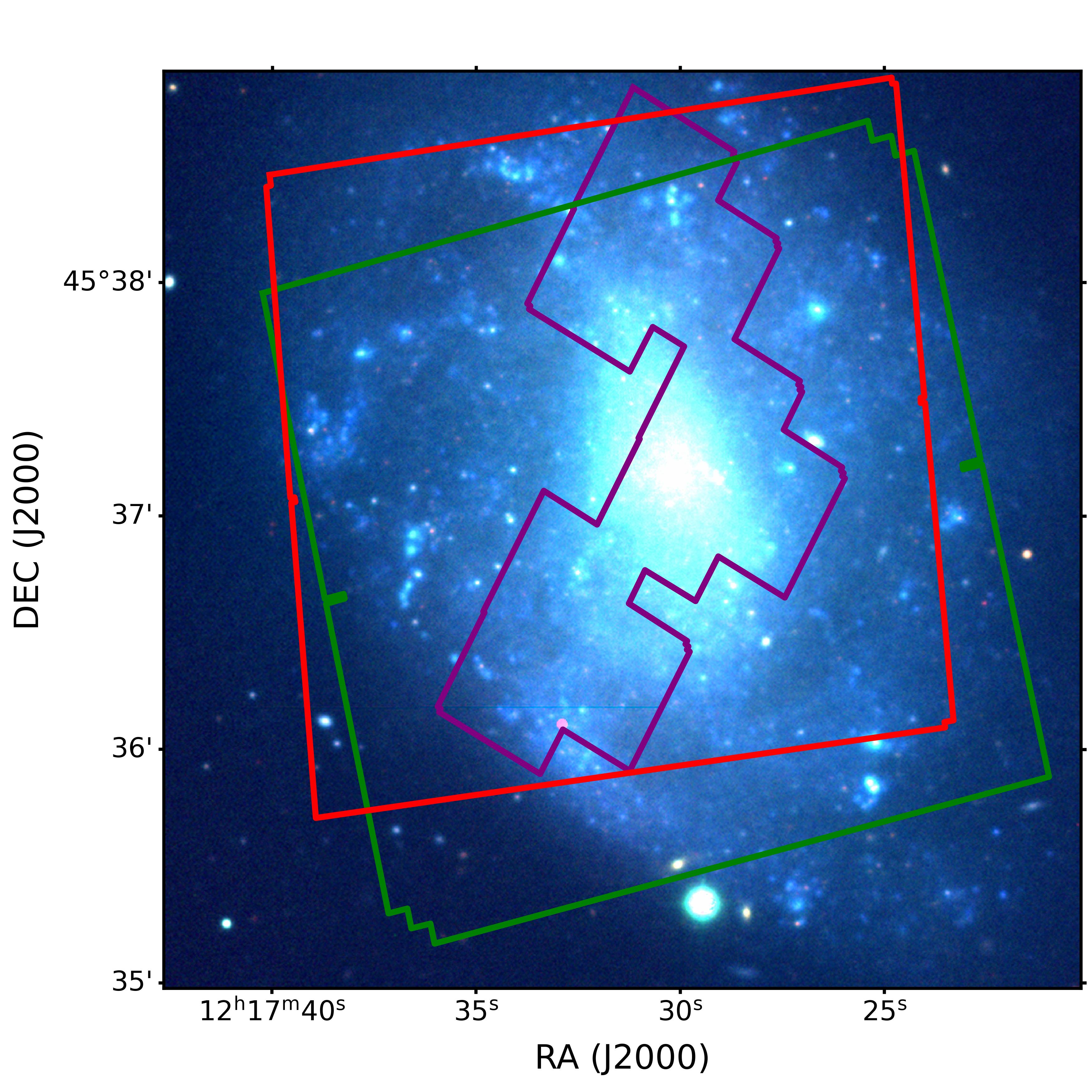}}
      \caption{Optical image of NGC $4242$ using the \textit{grz} filters from the Legacy Survey DR6 \citep{Dey2019} and the GULP (F$150$LP in purple, F$218$W in green) and LEGUS (F$275$W in red) mosaics.}
         \label{fig:FoV}
\end{figure}

The images of NGC $4242$ for our analysis were obtained using the Hubble Space Telescope's (HST) Solar Blind Channel of the Advanced Camera for Surveys F$150$LP filter (FUV) and the UVIS channel of the Wide Field Camera $3$ (UVIS/WFC$3$) filter F$218$W (MUV) from the Galaxy UV Legacy Project (GULP, PI Sabbi, GO-16316). Moreover, we used UVIS/F$275$W (NUV), F$336$W (U), F$438$W (B), F$555$W (V) and F$814$W (I) images from the HST Legacy ExtraGalactic UV Survey (LEGUS, PI  Calzetti, GO-13364, \citealt{Calzetti2015}) to extend the photometric coverage from the far-ultraviolet to the optical wavelengths. Figure~\ref{fig:FoV} shows the FUV, MUV and NUV mosaics used in this paper.

While the UV filters can be used to detect young massive stars, star clusters, and OB associations, the wide photometric coverage helps us to better constrain the spectral energy distributions (SEDs) between $\sim1500\text{\AA}$ and $\sim8100\text{\AA}$ of these high-mass stars, resulting in better estimates of their age and mass. Furthermore, the UV-filters are key factors to quantify how strongly/efficiently ultraviolet light emitted by massive stars is absorbed by dust grains.

The observing dates, instruments, filters and exposure times are reported in Table \ref{tab:Table filters}. All the frames were aligned to Gaia Data Release $3$ \citep{Gaia3} and drizzled to a common pixel scale of $0.04$"/px (the native WFC$3$ pixel size), using the F$438$W frame as reference. 

Fluxes for point-like sources in the F$275$W, F$336$W, F$438$W (reference frame), F$555$W and F$814$W filters were measured via PSF-fitting using the WFC$3$ module of the photometry package \texttt{DOLPHOT} version $2.0$ \citep{Dolphin2000,Dolphin2016}
The list of stars detected by \texttt{DOLPHOT} in the F$275$W filter was then used as input to perform aperture photometry in the filters F$150$LP and F$218$W using the Astropy package \texttt{Photutils} (Sabbi et al. in prep.).
The aperture radius around each object is $3$ pixel, and an annulus with an inner radius $12$ pixels and a width of $2$ pixels is used for sky subtraction.
All the magnitudes are in the Vegamag system.

\subsection{The stellar photometric catalog}\label{sec:Photometric catalog}

\begin{figure}
   \centering
   \resizebox{\hsize}{!}{\includegraphics{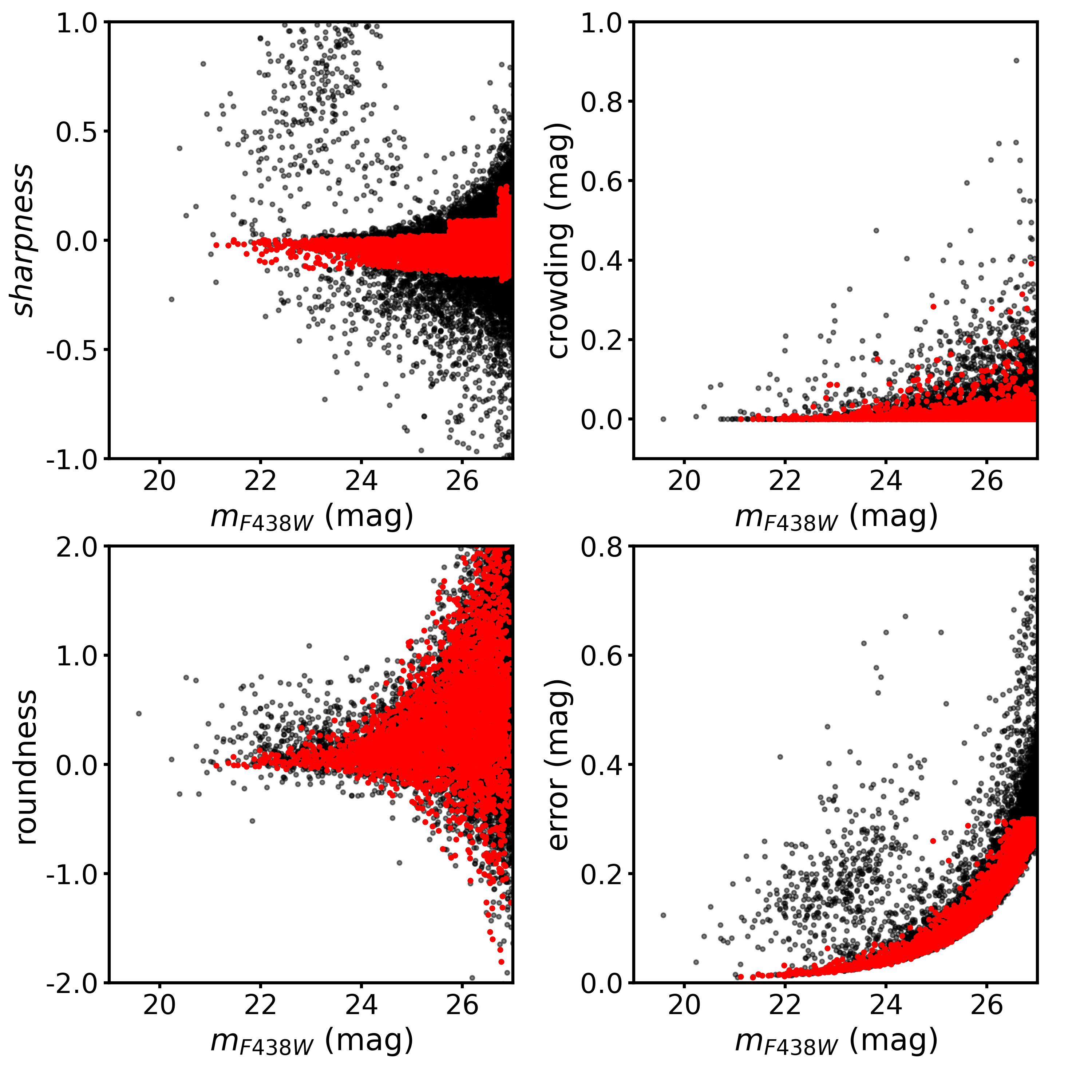}}
      \caption{Quality of the PSF photometry for the sources detected by LEGUS in the F$438$W (B) filter. In each panel, black dots refer to the full sample, while red points indicate sources that passed the full selection process as described in Section~\ref{sec:Photometric catalog}.}
         \label{fig:PSF}
\end{figure}

To correctly study the properties of massive stars, we need to remove as many extended objects or spurious detections as possible. We did this in two steps. First, we applied selection criteria on the photometric error, sharpness, and roundness of objects observed in the F$438$W and F$555$W filters. All sources with a photometric error larger than $0.3$ mag and $|\text{roundness}|$\textgreater$2$ were removed. A further constraint was put on the sharpness: objects whose sharpness deviates more than $1\sigma$ from the $50th$-percentile value per magnitude bin were discarded. Figure~\ref{fig:PSF} shows how the quality of the photometry changes as a function of the magnitude in the B filter. The black points represent the full catalog while red points show the sources that passed the full selection process. 

As a second step, out of the sources that passed the criteria above, we considered only those with a flux detection in the F$275$W filter. Stars that were observed in one or both FUV and MUV filters were selected only if their photometric errors in these filters are smaller than $0.3$ mag. At the end of the selection phase, we have: sources that are observed only in LEGUS filters (from the NUV to the I band), stars that are observed in all the filters (from the FUV to the I band), and sources with only one GULP flux. In fact, for this galaxy, GULP and LEGUS observations do not completely overlap and we did not want to miss any massive star candidates (see Figure~\ref{fig:FoV}).

\subsection{Star cluster and OB-association catalog}\label{sec:Star cluster catalog}

The LEGUS collaboration identified candidate star clusters and OB associations in the star forming galaxies observed by the survey \citep{Adamo2017,Grasha2017}. We summarize their procedure here.

An automatic pipeline extracted candidates and computed the concentration index (CI). The concentration index is defined as the magnitude difference between apertures at radii of $1$~pixel and $3$~pixels. A sample of isolated and bright stars was used to set the threshold CI value between stars and candidate star clusters. For NGC $4242$, the threshold value was $1.3$~mag for the WFC$3$ F$555$W filter. The CI cut allows the user to remove a large amount of stars from the catalog. However, compact and unresolved candidate star clusters whose effective radius is $1$~pc or smaller, can be potentially removed by the CI cut. Multi-band aperture photometry and averaged aperture corrections were performed for each target whose CI is larger than the threshold value. In order to remove possible outliers, the sources from the automatic catalog were visually inspected. All the candidates that were observed in at least four bands, with a photometric error smaller than $0.3$~mag and whose F$555$W absolute magnitude is brighter than $-6$~mag were visually inspected. During the visual inspection, sources that were clearly clusters and were missed during the automatic procedure were added to the catalog. The visually inspected sources were then divided in four classes. Class $1$ objects are centrally concentrated objects whose full width at half maximum (FWHM) is more extended than that of a star. Class $2$ objects have a less symmetric distribution of light while Class $3$ objects are characterized by a multi-peak distribution of their light. Finally, Class $4$ objects are potentially single stars or artifacts.

\begin{figure}
   \centering
   \resizebox{\hsize}{!}{\includegraphics{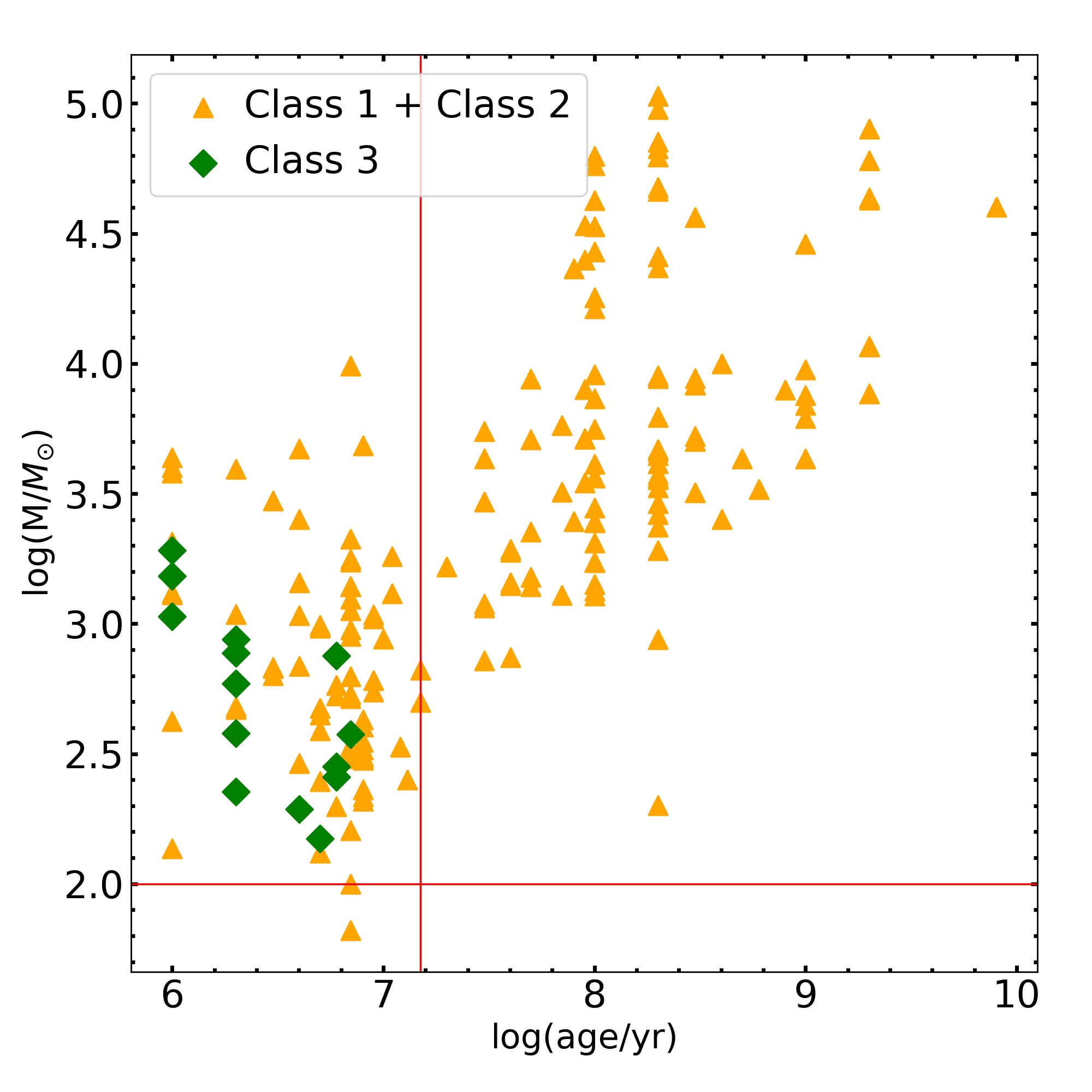}}
      \caption{Mass and age distribution of clusters/OB associations identified by the LEGUS collaboration using the theoretical models described in Section~\ref{sec:Star cluster catalog}. The red lines correspond to the $100 M_{\odot}$ cut-off and the $15$~Myr cut-off used in Section~\ref{sec:Isolation analysis}.}
         \label{fig:Clusters}
\end{figure}

The physical properties of the candidates (such as age, mass and dust extinction) were obtained using SED fitting techniques described in \citet{Adamo2017}. In this work, we used the catalog based on Yggdrasil SSP models \citep{Zackrisson2011} with a Kroupa initial mass function (IMF) \citep{Kroupa2001}, nebular continuum+emission lines with a covering factor $0.5$, metallicity Z=$0.02$, Padova isochrones \citep{Marigo2008, Girardi2010, Tang2014,Chen2015} and  Milky Way extinction law \citep{Cardelli1989}. Figure~\ref{fig:Clusters} shows the mass and age distribution of Class $1$ + Class $2$ and Class $3$ objects. In particular, it shows that a large fraction of clusters at young ages ($\leq 15$~Myr) are low-mass. This trend will be used in Section~\ref{sec:Isolation analysis} to refine the cluster sample.

\subsection{H$\alpha$ and infrared images}\label{sec:pure images}

In order to explore the environment around our massive star candidates, we used auxiliary H$\alpha$ and infrared images. In particular, we used the already stellar-continuum subtracted H$\alpha$ image taken by \citet{Kennicutt2008} with the Bok Telescope to identify the ionized regions around massive stars.

For the dust distribution in NGC $4242$, we used the infrared maps obtained by \citet{Dale2009} using the Spitzer Telescope: the Infrared Array Camera (IRAC) $8.0~\mu\text{m}$ image provides the information on the polycyclic aromatic hydrocarbon (PAH) molecules, while the Multiband Imaging Photometer (MIPS) $24.0~\mu\text{m}$ maps show the hot dust. We removed the star emission to get the pure PAH and hot dust emission using the $3.6~\mu\text{m}$ as continuum. We computed the stellar PSF of the $3.6, 8.0$ and $24.0~\mu\text{m}$ images. The pixel scale for the $3.6$ and $8.0~\mu\text{m}$ data is $0.75$"/px, while it is $1.5$"/px for the $24.0~\mu\text{m}$ data. We convolved the $3.6~\mu\text{m}$ image with the $8.0~\mu\text{m}$ PSF and the $8~\mu\text{m}$ image with the $3.6~\mu\text{m}$ PSF using the \texttt{IRAF} function \texttt{convolve}. In order to account for the different pixel scale of the $24.0~\mu\text{m}$ image, we first magnified the $3.6~\mu\text{m}$ image and PSF using the \texttt{magnify} function in \texttt{IRAF}, and then convolved it with the $24~\mu\text{m}$ image and PSF. The convolved images were then subtracted using the equation presented by \citet{Helou2004} and \citet{Dale2009}:
\begin{equation}
    \nu f_{\nu}~(8.0~\mu \text{m})_{dust} = \nu f_{\nu}~(8.0~\mu \text{m}) - \eta^{8*}~\nu f_{\nu}~(3.6~\mu \text{m}),
\end{equation}
and similarly for the $24.0~\mu\text{m}$ image, where $\nu$ is the frequency, $f_{\nu}$ is the specific flux density, $\eta^{8*}=0.232 \times 3.6/8.0$ for the $8.0~\mu\text{m}$ image and $\eta^{24*}=0.032 \times 3.6/24.0$ for the $24.0~\mu\text{m}$ map. We will refer to these stellar continuum-subtracted images as "pure" H$\alpha$, $8.0~\mu\text{m}$ and $24.0~\mu\text{m}$ images.

\section{Massive stars identification and characterization}\label{sec:Massive stars section}

\begin{figure*}
\centering
   \includegraphics[width=17cm]{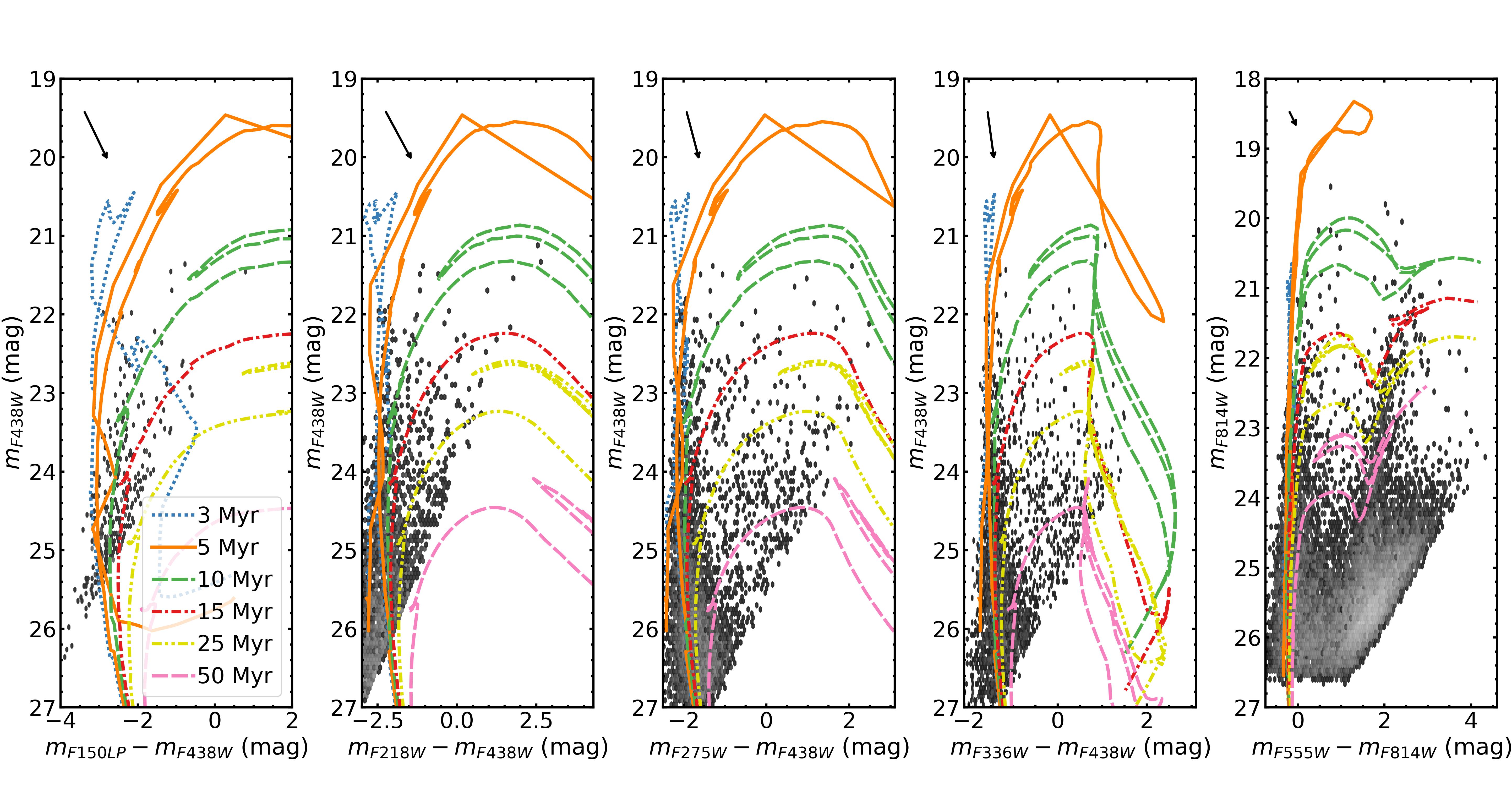}
     \caption{Hess diagrams for different combinations of filters. Padova isochrones for different ages are overplotted, using $Z=0.02$, $E(B-V)=0.05$ mag, distance modulus $\mu=28.61$ mag \citep{Sabbi2018} and the extinction coefficients from the \citet{Gordon2016} model with $R_V=3.1$ and $f_A=1.0$. The black arrow indicates the reddening vector using $A_V=0.5$~mag.}
     \label{fig:Extinction}
\end{figure*}

\begin{figure*}
\centering
   \includegraphics[width=17cm]{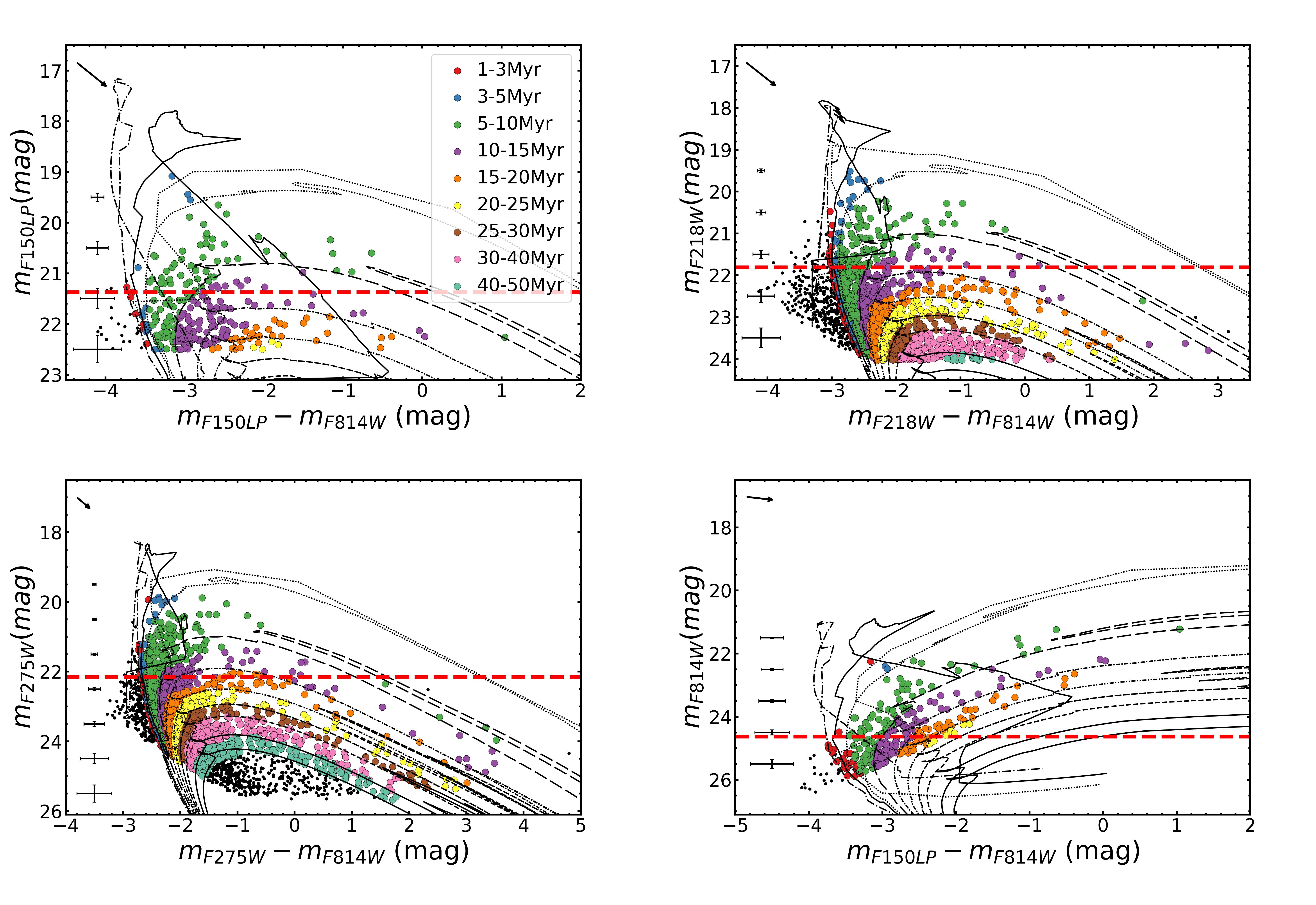}
     \caption{CMDs used to differentiate stars according to age. Padova isochrones for ages $1,3,5,10,15,20,25,30,40,50$ Myr, metallicity Z$=0.02$, distance modulus $\mu=28.61$ mag and intrinsic E(B-V)$=0.05$ mag are superimposed. The extinction coefficients come from \citet{Gordon2016} model with $R_V=3.1$ and $f_A = 1.0$. Stars are color-coded by the age determined by their positions between consecutive isochrones in each CMD. The black points are stars that do not lie in between consecutive isochrones. The dashed red line shows the luminosity of a main sequence star with $M_{ini}=15 M_{\odot}$ and an age of $10$~Myr. The black arrow indicates the reddening vector using $A_V=0.1$~mag.}
     \label{fig:CMDs}
\end{figure*}

Color-magnitude diagrams (CMDs) are useful tools for interpreting the resolved stellar component in the host galaxy. 
Massive stars show the peak of their emission at ultraviolet wavelengths, making them bright objects in the UV filters of the GULP and LEGUS surveys.
However, the dust component of the host galaxy absorbs a large fraction of the ultraviolet light, re-emitting it at longer wavelengths. Moreover, different empirical extinction laws for different environments have been found.
Thus, it is crucial to determine the correct extinction law for NGC~$4242$ in order to interpret the properties of its massive stars.

\subsection{Extinction Law}\label{sec:Extinction Law}

How extinction changes as a function of the wavelength was measured for a number of galaxies including the Milky Way \citep{Cardelli1989,Fitzpatrick1999} and the Magellanic Clouds \citep{Nandy&Morgan1978,Gordon2003,MaizApellaniz&Rubio2012}. While the MW extinction law exhibits a bump in the absorption of photons at a wavelength of $2175$\text{\AA} (the so-called UV-bump), the SMC extinction law does not.

We first removed the Galactic foreground extinction using \citeauthor{Cardelli1989}'s \citeyearpar{Cardelli1989} extinction law and a value of $A_V=0.033$~mag \citep{Schlafly2011}. We then used the mixture model from \citet{Gordon2016} with a Milky Way and a Small Magellanic Cloud component to see how the extinction affects the photometry of our data. This model depends on the total-to-selective extinction ratio, $R_V$, and $f_A$ which regulates the amount of extinction due solely to the Milky Way component \citep[see][]{Gordon2016}. We chose the $R_V$ and $f_A$ parameters by visually inspecting the V-I and UV Hess diagrams in two steps. First, we varied the $R_V$ value to match the V-I Hess diagram. Then, keeping $R_V$ fixed, we fine-tuned the $f_A$ parameter trying to match the blue edge in the FUV-B and MUV-B Hess diagrams. We found that the distribution of stars in different Hess diagrams (from the FUV to the I band) is well described by Padova isochrones \citep{Marigo2008,Girardi2010,Tang2014,Chen2015} using a true distance modulus $\mu=28.61$ mag, an intrinsic extinction of $E(B-V)=0.05$ mag and a metallicity $Z=0.02$ \citep{Sabbi2018} for the pair $(R_V,f_A)=(3.1,1.0)$. This corresponds to the Milky Way extinction law (\citealt{Fitzpatrick1999} with $R_V=3.1$) for NGC $4242$. 
Figure~\ref{fig:Extinction} shows the aforementioned Hess diagrams with overplotted Padova isochrones corrected for extinction using $R_V=3.1$ and $f_A=1.0$.

\subsection{Age Separation}\label{sec:Age separation}

We decided to differentiate stars according to their age estimated via the F$150$LP vs. F$150$LP-F$814$W, F$814$W vs. F$150$LP-F$814$W, F$218$W vs. F$218$W-F$814$W and F$275$W vs. F$275$W-F$814$W  CMDs. Figure~\ref{fig:CMDs} shows the four different CMDs using Padova isochrones and the extinction parameters from Section~\ref{sec:Extinction Law}. Stars were divided in age ranges according to their position within two consecutive isochrones.

We are only interested in massive and young objects, so we removed all the objects with an initial mass smaller than $15 M_{\sun}$ from each CMD. Using Padova isochrones and a foreground Milky Way extinction $A_V=0.033$ mag \citep{Schlafly2011}, we removed all the objects fainter than $24.7$, $21.4$, $22.2$ and $21.8$ mag for the F$814$W, F$150$LP, F$275$W and F$218$W filters, respectively. \citet{Schneider2018} analyzed $173$ O stars in $30$ Doradus finding an average maximum age of about $10$~Myr. We therefore decided to consider objects in the age range $1-10$~Myr. Stars that are bluer than the $1$~Myr isochrone (possibly due to photometric errors, uncertainty on the metallicity, the extinction law, and theoretical models) were also taken into account. 
At this point, stars that are younger than $10$~Myr and brighter than the magnitude cut in at least one of these CMDs are considered. We selected $242$ objects from the four CMDs. Assuming that these objects are single non-rotating stars, we estimated their initial mass ($M_{ini}$), age and intrinsic extinction ($A_{V}$) based on their photometry minimizing the difference between observed and theoretical magnitudes.
Let \textbf{y}$_{obs}$ and $\boldsymbol{\sigma}$ be the vectors of the observed magnitudes and photometric errors, and \textbf{y}$_{th}$ the vector of synthetic reddened magnitudes derived from Padova isochrones. In the theoretical models, the age varies between $1$ and $10$~Myr with steps of $0.2$~Myr, $M_{ini}$ varies between $0.08$ and $350 M_{\odot}$ and $A_V$ ranges from $0.0$ to $1.0$ mag with steps of $0.01$ mag. We computed the $\chi^2$ for the single model as
\begin{equation}
    \chi^{2} = \sum^{n_{filters}}_{i=1} \frac{\left( y_{obs,i}- y_{th,i}\right)^2}{\sigma_{i}^2}
\end{equation}
and the sum is done over the number of filters.
The minimum $\chi^2$ gave the best (logt,$M_{ini}$,$A_V$) triplet. To estimate the error on each parameter, we perturbed the observed photometry of each star using a multivariate normal distribution, $\mathcal{N}(\mu=\boldsymbol{y}_{obs},\,\sigma^{2}=\boldsymbol{\sigma}^{2})$, centered on the observed photometry and with the photometric errors as its $\sigma$. We ran $100$ Monte Carlo iterations for each star and took the standard deviations of the $100$ best triplets as the corresponding errors of each parameter.

The $\chi^2$ confirmed that $234$ sources are younger than $10$~Myr and with an initial mass larger than $15 M_{\odot}$: $100$ stars have fluxes available in all filters, $110$ with only one GULP flux available and the rest ($24$) with only LEGUS fluxes (see Section~\ref{sec:Photometric catalog} and Figure~\ref{fig:FoV}).

\begin{figure*}
   \centering
   \includegraphics[width=17cm]{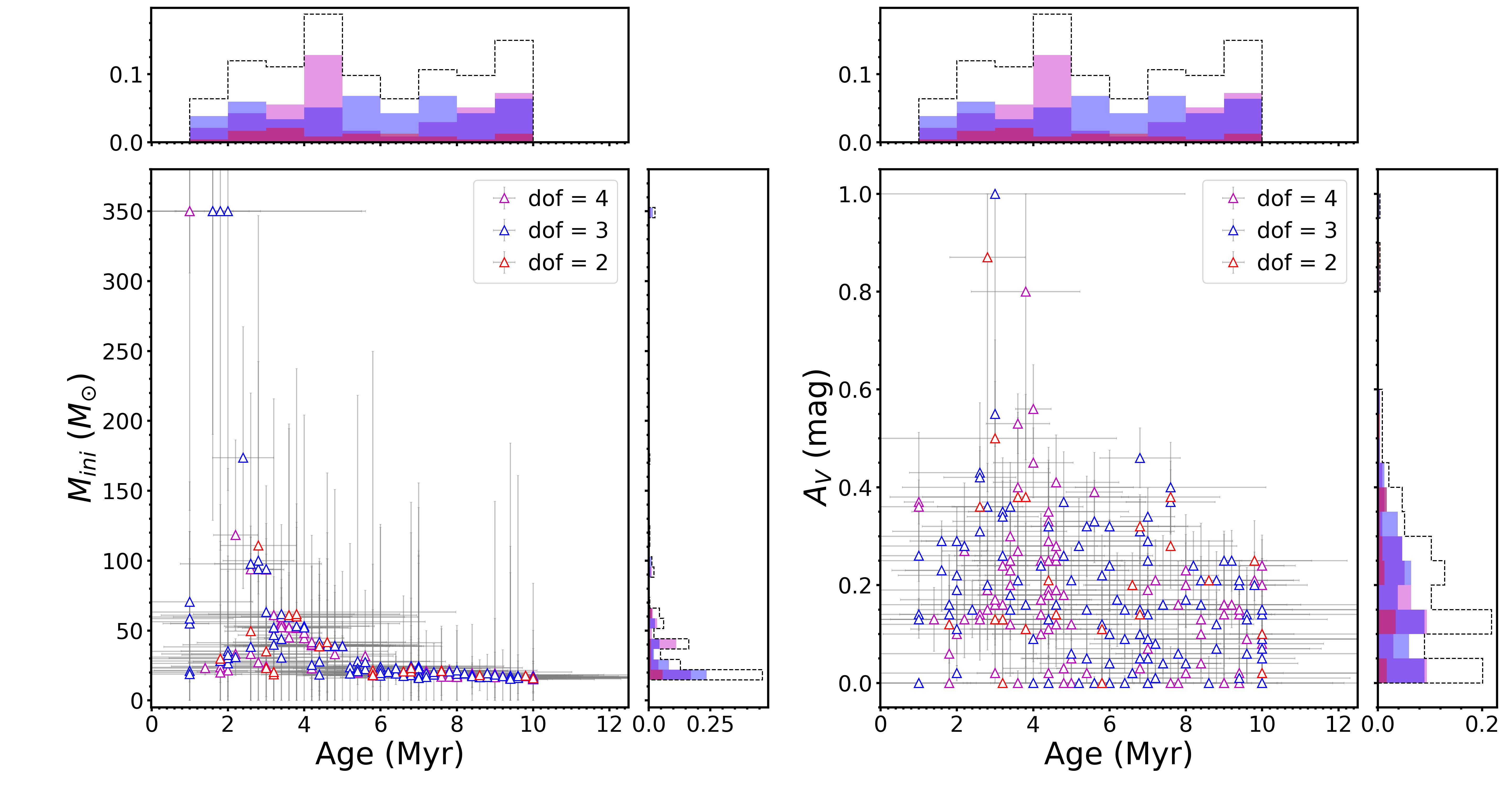}
      \caption{Best values for age, $A_V$ and $M_{ini}$ and their histogram distributions using Padova isochrones with a fixed metallicity $Z=0.02$, an age range of $1-10$~Myr and a varying intrinsic extinction $A_V \in [0.0,1.0]$ based on the extinction coefficients from \citet{Gordon2016} for $R_V=3.1$ and $f_A=1.0$ (Section \ref{sec:Extinction Law}). The targets are divided in three groups based on the degrees of freedom, which depend on the number of available filters as explained in Section~\ref{sec:Photometric catalog} ($\text{dof}=4$ for complete wavelength coverage, $\text{dof}=3$ for only one GULP flux and $\text{dof}=2$ for detection only in between the NUV and the I bands). The histograms are normalized to the total number of sources. The black dashed histogram shows the total frequency.}
         \label{fig:Best}
\end{figure*}

Figure~\ref{fig:Best} shows the best values obtained for $A_V$, age and $M_{ini}$ color-coded by the degree of freedom ($\text{dof}=4$ for complete wavelength coverage, $\text{dof}=3$ for only one GULP flux and $\text{dof}=2$ for detection only in between the NUV and the I bands). The histograms show that the age and $A_V$ of our selected stars are distributed throughout the age range and the $0.0$ to $1.0$ mag interval, with a mean age of $5.7$~Myr and an average intrinsic extinction of $A_V=0.18$ mag, which is in agreement with \citet{Sabbi2018}. The majority of our stars have $M_{ini}$ between $15$ and $100M_{\odot}$, with some of them exceeding $M_{ini}=150M_{\odot}$ and $A_V=0.4$ mag. A possible explanation for the latter is the presence of a star cluster around our targets or other unresolved stars that can contaminate the photometry. In fact, some of our targets are close to star clusters identified by LEGUS \citep[see][]{Adamo2017}, making the $\chi^2$ higher and the parameter estimate less precise.

In order to test how our derived stellar ages, initial masses and dust extinctions depend on the metallicity of the isochrone and on the extinction law assumed for NGC $4242$, we fitted the observed photometry using three different setups: Padova isochrones with SMC metallicity reddened with MW extinction law, Padova isochrones with solar metallicity and SMC extinction law ($R_V=2.74$ \citealt{Gordon2003}), and Padova isochrones with SMC metallicity and SMC extinction law. We found that, at fixed metallicity, the average age of our sample becomes younger by $1$~Myr and the average extinction slightly increases by $0.01$ mag, while at fixed extinction law, the population becomes older by $1$~Myr and the extinction increases not more than $0.02$ mag. The results are described in Appendix \ref{appendix}.

\subsection{Spatial distribution of the targets}\label{sec:spatial distribution}
\begin{figure}[t!]
  \resizebox{\hsize}{!}{\includegraphics{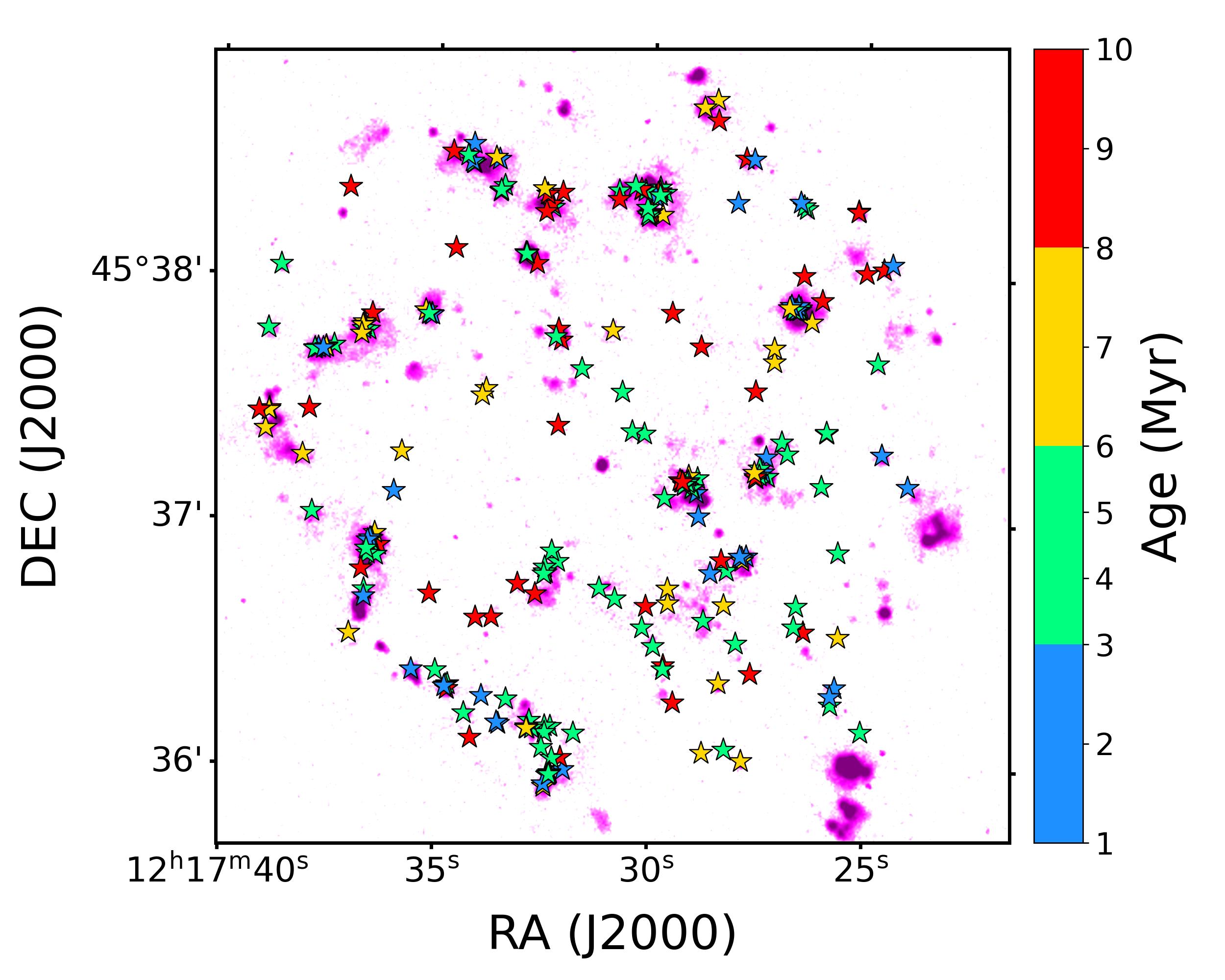}}
  \caption{Spatial distribution of the selected targets grouped in four different age bins. The underlying pure H$\alpha$ emission (in magenta) is taken by \citet{Kennicutt2008} with the Bok Telescope.}
  \label{fig:Halpha}
\end{figure}

We expect to find massive and young objects along the spiral arms of NGC $4242$. Figure~\ref{fig:Halpha} shows the spatial distribution of the targets grouped in four different age bins. We use the stellar continuum subtracted H$\alpha$ image taken by \citet{Kennicutt2008} with the Bok Telescope to connect young targets with H$\alpha$ emission (drawn in magenta). The comparison between Figure \ref{fig:FoV} and \ref{fig:Halpha} shows that most of the targets are distributed along the faint spiral arms of NGC $4242$, as delineated by the H$\alpha$ emission in Figure~\ref{fig:Halpha}.

\begin{figure*}
     \centering
     \includegraphics[width=17cm]{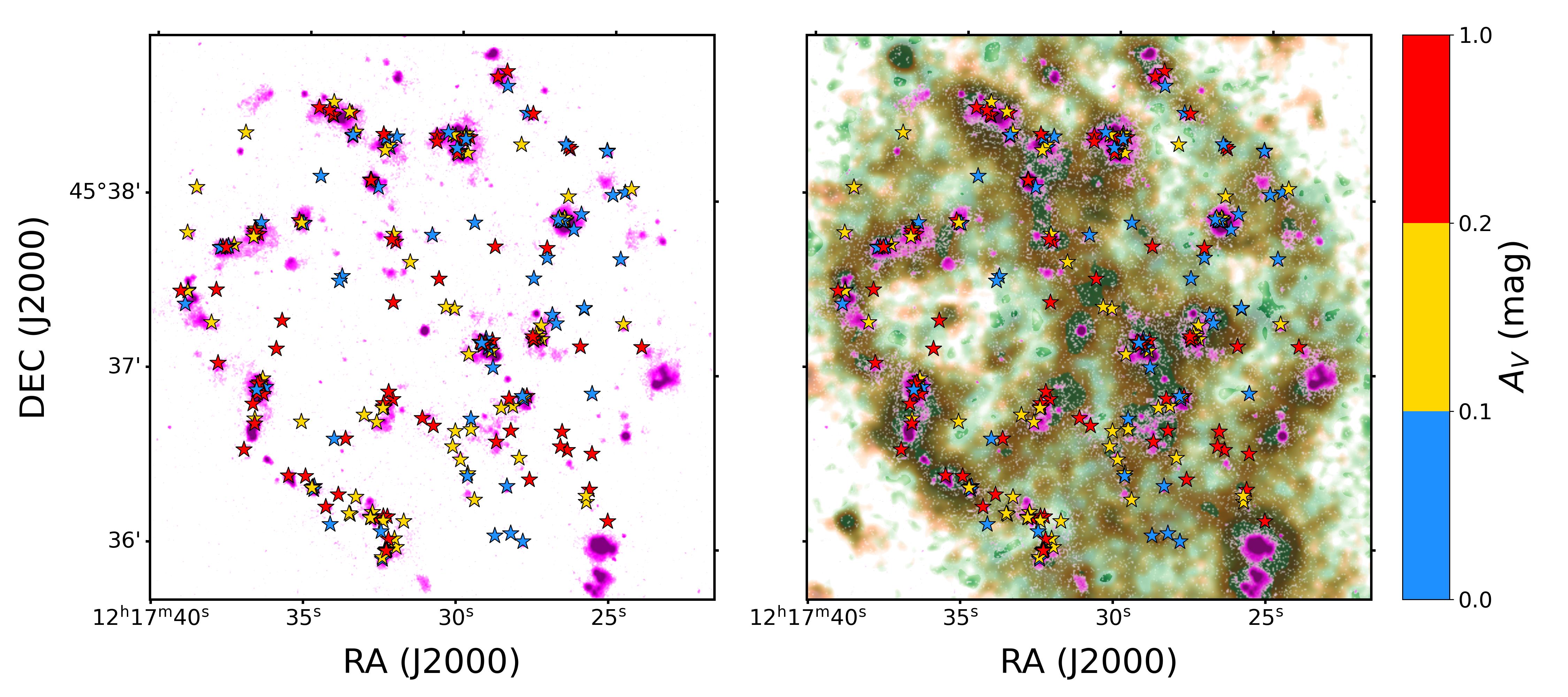}
     \caption{Spatial distribution of the targets color-coded by the best value of their intrinsic extinction $A_V$. On the left panel the pure H$\alpha$ emission is drawn in magenta. On the right panel, the underlying pure H$\alpha$ is in magenta, the pure $8.0~\mu\text{m}$ in green (PAH emission), and the pure $24.0~\mu\text{m}$ in orange (hot dust emission).}
     \label{fig:Dust}
\end{figure*}

In most of the cases, our targets are located in H$\alpha$ emission regions. The left panel of Figure~\ref{fig:Dust} shows the spatial distribution of the targets color-coded by the best values for their reddening. We note that some targets need a very small reddening even if they coincide with H$\alpha$ emission. We suggest that these targets are not in the H$\alpha$ emission region, but they are in the foreground and the H$\alpha$ is in the background within the galaxy NGC $4242$. On the other hand, some targets with a high reddening are not associated with H$\alpha$ emission. It is possible that stars with a high value for the reddening are behind dust lanes. The right panel of Figure~\ref{fig:Dust} shows a composite image of the pure H$\alpha$ emission (in magenta), the pure $8.0~\mu\text{m}$ (in green) and the pure $24.0~\mu\text{m}$ (in orange). The dust images were taken by the Spitzer Space Telescope \citep{Dale2009} and the stellar continuum subtraction is described in Section~\ref{sec:pure images}.

\section{Isolation analysis}\label{sec:Isolation analysis}

\subsection{Isolation with respect to star clusters or OB associations}

\begin{figure*}
\resizebox{\hsize}{!}{\includegraphics{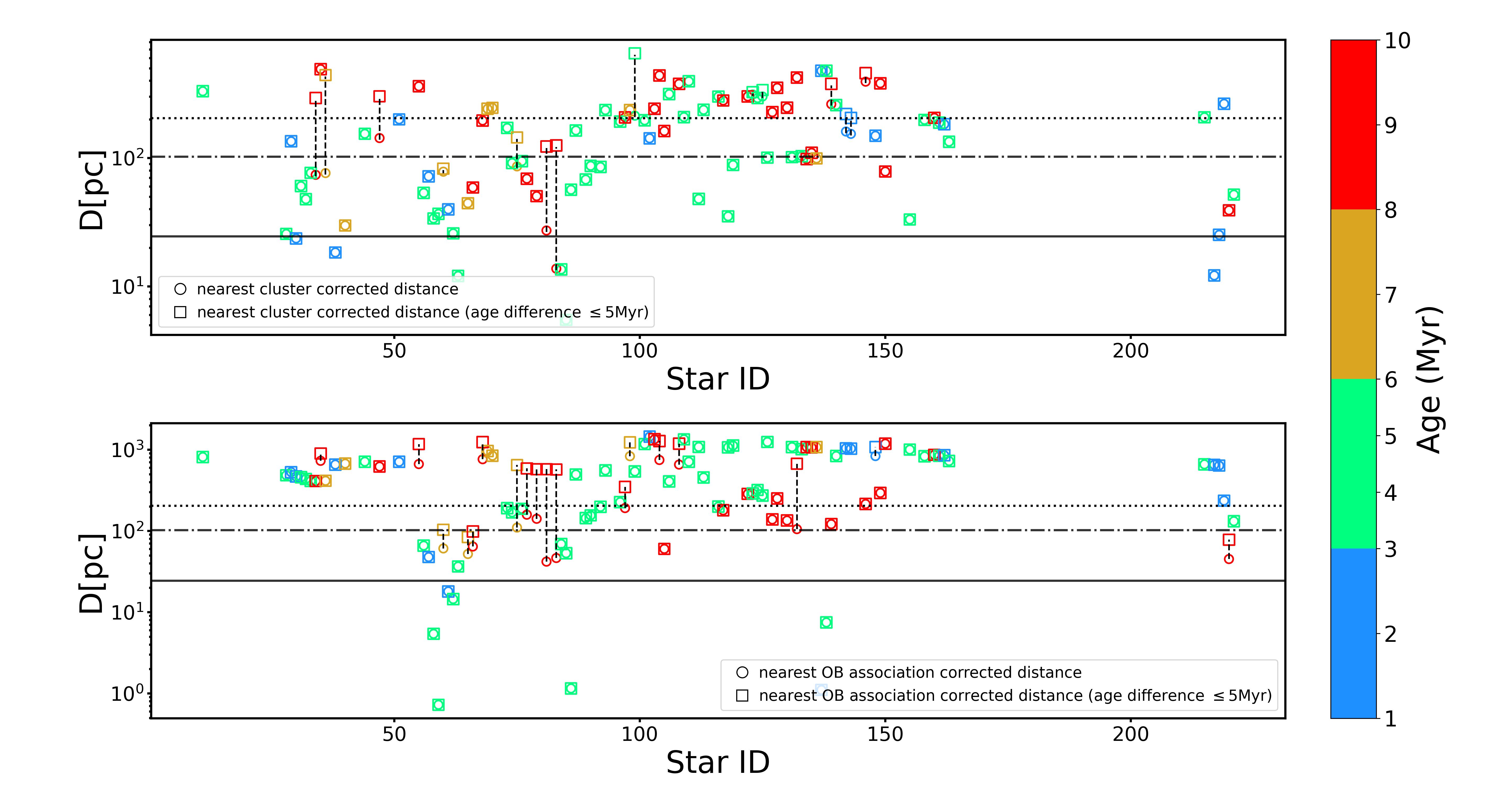}}
      \caption{De-projected distances between stars with $\text{dof}=4$ and star clusters (top panel) or OB associations (bottom panel). On the x axis the star ID from the sample is shown. The circle symbols show the distance of the closest cluster from each target (Method $1$), while the squares show the closest cluster with an age similar to that of each star (Method $2$). The black dashed vertical lines connect the distances from Method $1$ and Method $2$ when they find two different stellar groups. The symbols are color-coded accordingly to the derived age of the stars. The gray horizontal lines show the maximum tangential distances that a star can travel in $1$~Myr with a constant velocity of $24~\text{km}~\text{s}^{-1}$, $100~\text{km}~\text{s}^{-1}$ and $200~\text{km}~\text{s}^{-1}$. These distances are $24$, $102$ and $204$~pc (horizontal solid, dot-dashed and dotted lines, respectively). At a distance of $5.3$ Mpc, $1$ pc is $0.039$ arcsec.}
         \label{fig:Distance_dof4}
\end{figure*}

\begin{figure*}
   \centering
   \includegraphics[width=17cm]{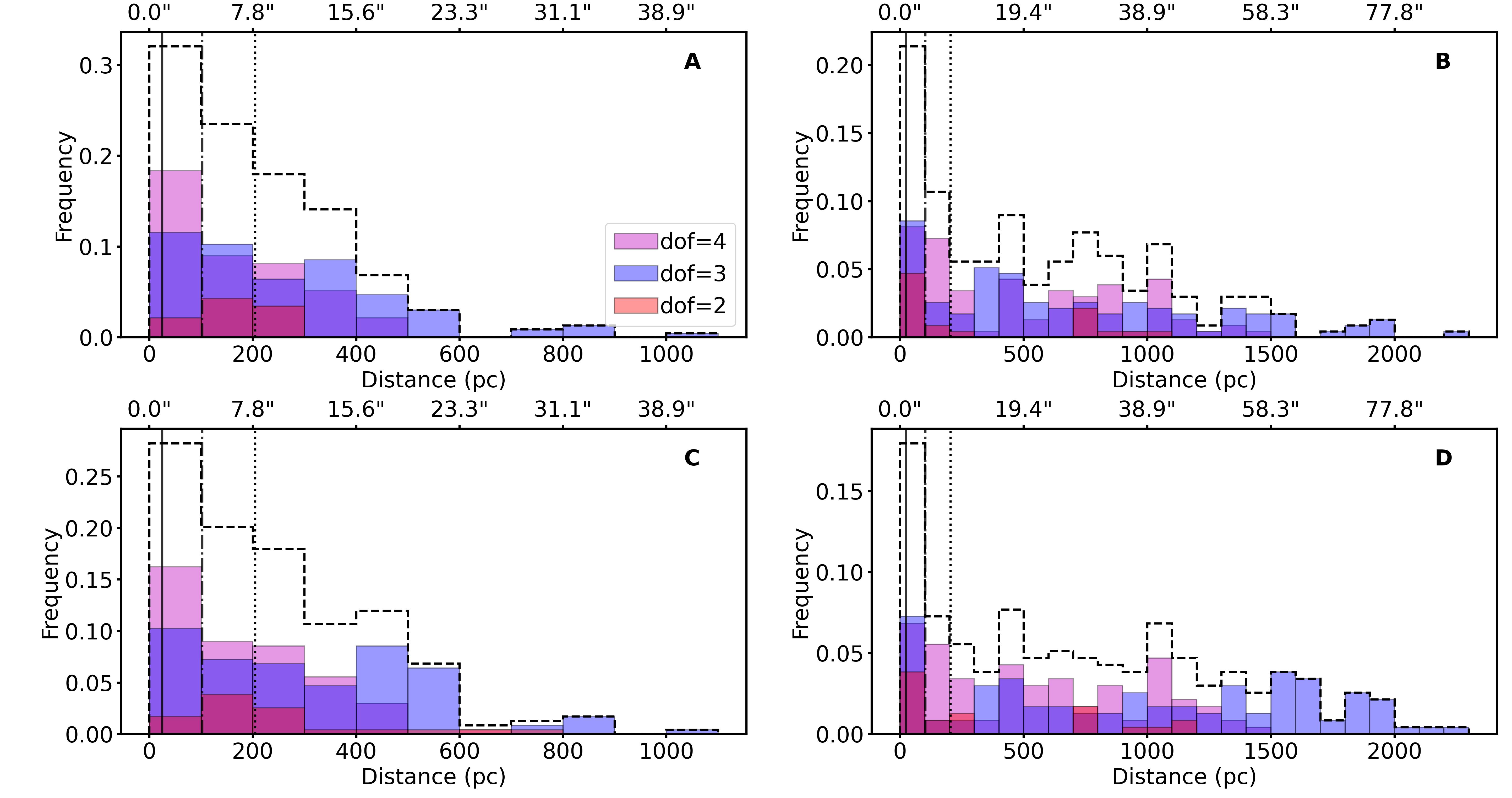}
      \caption{Distance distributions: panels A and B show the simple projected $2$D distance between our stars and the nearest star cluster and OB association, respectively; panels C and D show the distances from the nearest star cluster or OB associations differing from estimated stellar ages by $5$~Myr at most. The sample is color-coded accordingly to the degree of freedom and each histogram is normalized to the total number of stars in the sample. The dashed black histogram shows the total frequency. All bins have equal size of $100$~pc. The black vertical lines show the distances that a star can travel in $1$~Myr with a constant velocity of $24~\text{km}~\text{s}^{-1}$, $100~\text{km}~\text{s}^{-1}$ and $200~\text{km}~\text{s}^{-1}$. These distances are $24$, $102$ and $204$~pc.}
         \label{fig:Histo}
\end{figure*}

Even if massive stars are short-lived, those in the field are frequently considered to be runaways or walkaways from known star clusters or OB associations \citep{Blaauw1961,Poveda1967,Gies1987,Hoogerwerf2001,DorigoJones2020}. Thus, we looked at the spatial distribution of our targets and the visually inspected young star clusters (age $\leq15$~Myr) from \citet{Adamo2017} (see Section \ref{sec:Star cluster catalog}) to see how many stars from our sample could potentially be runaways from star clusters. We note here that while in the MW and the SMC, the runaway or walkaway nature of a star can be tested via spectroscopic radial velocities and astrometric tangential velocities, this is not feasible with the existing instrumentation in a galaxy at the distance of NGC $4242$. 

The clusters can be divided in three classes, depending on their PSF shape: spherically symmetric, asymmetric, or multipeak PSF (see Section~\ref{sec:Star cluster catalog}). In this work, we grouped together symmetric and asymmetric objects and we will refer to these objects as star clusters, while multipeak PSF objects will be considered OB associations due to their larger extensions.
We decided to consider only objects more massive than $100 M_{\odot}$. In fact, using Gaia data, \citet{Hunt2023} were able to identify Galactic open clusters as massive as $\sim 100M_{\odot}$ containing few massive stars as massive as $50M_{\odot}$.
In NGC~$4242$ we have $74$ young ($\leq 15$~Myr) star clusters and only $14$ OB associations, with a median separation of $\sim180$~pc.

Runaway OB stars can be produced via two main mechanisms: the dynamical ejection scenario and the binary supernova scenario. In the former, gravitational interactions in dense clusters could eject a massive star \citep{Blaauw1961,Poveda1967}, while in the latter the supernova explosion could eject the companion star \citep{Renzo2019,Blaauw1961}. Recently, \citet{Polak2024} have analyzed the production of runaways through the subclusters ejection mechanism, where stars in a subcluster can be ejected during the tidal disruption of the parent subcluster falling into the deeper gravitational potential of the proto cluster. Runaway stars are traditionally defined to have a $3$D peculiar motion of at least $30$~km~$\text{s}^{-1}$ \citep{Gies1987,Hoogerwerf2001}, which corresponds to a transverse $2$D minimum velocity of $24~\text{km}~\text{s}^{-1}$ assuming equal velocity components in space \citep{DorigoJones2020}. Fast runaways with a velocity of $200~\text{km}~\text{s}^{-1}$ in the SMC have been found by \citet{DorigoJones2020}, however, theoretical studies \citep{Perets2012} show that the runaway frequency is strongly reduced with increasing peculiar velocity. 
This is in agreement with \citet{DorigoJones2020}, who found that only $48$ stars out of $199$ runaway stars in the SMC have local velocities larger than $75~\text{km}~\text{s}^{-1}$, showing that the peak of the distribution of runaway stars is reached between $30$ and $75~\text{km}~\text{s}^{-1}$ with a steep drop for higher velocities. A steep decrease in the distribution of runaway stars was also found for the massive star cluster R$136$ in the Large Magellanic Cloud \citep{Stoop2024}, with only $8$ (out of $23$) O-type stars being runaways from R$136$ with tangential velocities larger than $50~\text{km}~\text{s}^{-1}$. Taking these velocities as references, a runaway star can travel tens to hundreds of parsecs in a few Myr. For example, \citet{Stoop2024} found that the population of O runaway stars has traveled a mean distance of $59$~pc from the center of R$136$, with this value reaching a value of $115$~pc considering also the runaways without a spectral classification. The theoretical studies on the BSS done by \citet{Renzo2019} revealed that a main sequence massive companion star travels on average $\sim 75$~pc, in agreement with the $50$~pc found by \citet{Eldridge2011}. However, both distributions of the maximum distance are characterized by a large scatter.

We looked at the closest star cluster or OB association from each star target, and addressed the question "How probable is it that our target belongs to that star cluster?". We de-projected the positions of stars and star clusters on the galaxy plane adopting an inclination of $i=40.5^\circ$ \citep{Calzetti2015} and then computed their $2$D projected distance. 

In order to take into account the velocity distribution found by \citet{DorigoJones2020}, we decided to take two threshold distances: $d_{thresh,1}=74$~pc and $d_{thresh,2}=204$~pc. The former corresponds to the distance traveled by a star in $1$~Myr with three times the minimum $2$D runaway velocity, while the latter is the distance that a fast runaway can travel in $1$~Myr at $200~\text{km}~\text{s}^{-1}$. These two threshold distances allowed us not only to identify runaway candidates with the former velocities and $1$~Myr travel time, but also runaways with a combination of velocities and travel times. For example, the $d_{thresh,2}$ can help identify runaway stars with the minimum $2$D runaway velocity that have traveled for $\sim8$~Myr.

\begin{table*}
\caption{Number and frequency of possible isolated stars with respect to the closest star clusters or OB associations.}
\label{tab:Frequency}
\centering
\begin{tabular}{ccccc}
\hline \hline
Sample & \multicolumn{2}{c}{Method $1$} & \multicolumn{2}{c}{ Method $2$}\\
\hline \hline
$d_{thresh,1}=74$pc & Number & Frequency & Number & Frequency\\
\hline
only clusters\\
dof=$4$ & $70$ & $0.299$ & $72$ & $0.308$ \\
dof=$3$ & $86$ & $0.368$ & $90$ & $0.385$\\
dof=$2$ & $19$ & $0.081$ & $20$ & $0.085$\\
whole sample & $175$ & $0.748$ & $182$ & $0.778$\\
\hline
only OB associations\\
dof=$4$ & $81$ & $0.346$ & $87$ & $0.372$\\
dof=$3$ & $92$ & $0.393$ & $94$ & $0.402$\\
dof=$2$ & $15$ & $0.064$ & $16$ & $0.068$\\
whole sample & $188$ & $0.803$ & $197$ & $0.842$\\
\hline
clusters + OB associations \\
dof=$4$ & $66$ & $0.282$ & $69$ & $0.295$\\
dof=$3$ & $80$ & $0.342$ & $85$ & $0.363$\\
dof=$2$ & $13$ & $0.056$ & $15$ & $0.064$\\
whole sample & $159$ & $0.679$ & $169$ & $0.722$\\
\hline \hline
$d_{thresh,2}=204$pc & Number & Frequency & Number & Frequency\\
\hline
only clusters \\
dof=$4$ & $35$ & $0.149$ & $40$ & $0.171$\\
dof=$3$ & $57$ & $0.244$ & $68$ & $0.291$\\
dof=$2$ & $9$ & $0.038$ & $11$ & $0.047$\\
whole sample & $101$ & $0.432$ & $119$ & $0.509$\\
\hline
only OB associations \\
dof=$4$ & $64$ & $0.274$ & $71$ & $0.303$\\
dof=$3$ & $84$ & $0.359$ & $91$ & $0.389$\\
dof=$2$ & $11$ & $0.047$ & $13$ & $0.056$\\
whole sample & $159$ & $0.679$ & $175$ & $0.748$\\
\hline
clusters + OB associations \\
dof=$4$ & $26$ & $0.111$ & $33$ & $0.141$\\
dof=$3$ & $49$ & $0.209$ & $62$ & $0.265$\\
dof=$2$ & $4$ & $0.017$ & $8$ & $0.034$\\
whole sample & $79$ & $0.338$ & $103$ & $0.440$\\
\hline
\end{tabular}
\tablefoot{Number and frequency of possible isolated stars (with respect to the closest star clusters or OB associations) given the threshold distance of $d_{thresh,1}=74$ pc and $d_{thresh,2}=204$ pc. Method $1$ takes into account only the de-projected positions of stars and clusters, while Method $2$ considers clusters whose age difference with the stars is $5$~Myr at most. The frequencies are normalized to the total sample of stars.}
\end{table*}

Every star that has a star cluster/OB association closer than the threshold distance can be considered as possible walkaway or runaway from that particular group, otherwise it is in isolation (thereafter Method $1$). Each circle in Figure~\ref{fig:Distance_dof4} shows the location of the closest star cluster/OB association for each star with dof$=4$. However, finding a star near to a star cluster does not imply that they are connected. We decided to look also at the age information: given the age of the star, we selected star clusters whose age differs by $5$~Myr at most from the age of the star, and then searched for the closest cluster (thereafter Method $2$). The $5$~Myr range takes into account the age spread of O-type stars found in Cygnus OB2 \citep{Berlanas2020} and in the massive star cluster R$136$ \citep{Schneider2018}. Moreover, it corresponds to the maximum error on the age of our stars computed in Section~\ref{sec:Age separation}. In fact, the mean error on the age of a stars from the isochrone fitting is $1.8$~ Myr, while the maximum error is $5$~Myr. We chose the latter as the maximum age difference between a stars and a cluster to create possible connections between them.
Each square in Figure~\ref{fig:Distance_dof4} shows where the closest star cluster/OB association is located for each star with a similar age. The stars are divided into age bins. 
The two methods for the distances in some cases found the same connection between star and clusters (no connecting dashed line), while there are some stars that appear to be close to a star cluster whose age difference is more than $5$~Myr, making their connections less likely true, and increasing their chance to be unrelated to those clusters.

Following the age bins in Figure~\ref{fig:Distance_dof4}, we computed the median separation between stars and their nearest cluster/OB association using Method 1 and 2. For the star clusters we found a positive correlation, with the median distance between young stars and star clusters increasing towards older age bins; however, this trend becomes weaker for Method $2$. On the other hand, for younger age bins, OB associations are found to be closer to young stars with the trend disappearing using Method 2. The different trend between clusters and OB associations is most likely due to the small number statistics of the latter.

In Figure~\ref{fig:Histo} we show how many connections between star and star clusters/OB associations we are able to find as a function of the projected distance. The figure shows that there is a population of stars that are found far away from known star clusters or OB associations, being potentially isolated. 
Table~\ref{tab:Frequency} summarizes the number of isolated objects given the method or the threshold value. Note that some of our stars could have traveled a longer distance (with a higher runaway velocity or for longer time), and may contaminate the isolated star sample. Conversely, stars in projected proximity to clusters or OB associations may be unrelated to them.

\subsection{Isolation with respect to other massive stars}

\begin{table*}
\caption{Number of possible isolated stars with respect to the closest star clusters, OB associations and massive stars.}
\label{tab:Isolation stars}
\centering
\begin{tabular}{ccccc}
\hline \hline
Sample & \multicolumn{2}{c}{Method $1$} & \multicolumn{2}{c}{Method $2$}\\
\hline
\ & Number & Frequency & Number & Frequency\\
\hline \hline
clusters + OB associations ($d_{thresh,1}$) \\
and stars ($d_{thresh,1}$) isolation\\
dof=$4$ & $31$ & $0.132$ & $31$ & $0.132$\\
dof=$3$ & $42$ & $0.179$ & $43$ & $0.184$\\
dof=$2$ & $6$ & $0.026$ & $7$ & $0.029$\\
whole sample & $79$ & $0.338$ & $81$ & $0.346$\\
\hline
clusters + OB associations ($d_{thresh,1}$) \\
and stars ($d_{thresh,2}$) isolation\\
dof=$4$ & $6$ & $0.026$ & $6$ & $0.026$\\
dof=$3$ & $20$ & $0.085$ & $20$ & $0.085$\\
dof=$2$ & $2$ & $0.009$ & $2$ & $0.009$\\
whole sample & $28$ & $0.120$ & $28$ & $0.120$\\
\hline
clusters + OB associations ($d_{thresh,2}$) \\
and stars ($d_{thresh,1}$) isolation\\
dof=$4$ & $15$ & $0.064$ & $17$ & $0.073$\\
dof=$3$ & $29$ & $0.124$ & $33$ & $0.141$\\
dof=$2$ & $2$ & $0.009$ & $5$ & $0.021$\\
whole sample & $46$ & $0.197$ & $55$ & $0.235$\\
\hline
clusters + OB associations ($d_{thresh,2}$) \\
and stars ($d_{thresh,2}$) isolation\\
dof=$4$ & $4$ & $0.017$ & $4$ & $0.017$\\
dof=$3$ & $17$ & $0.073$ & $18$ & $0.077$\\
dof=$2$ & $2$ & $0.009$ & $2$ & $0.009$\\
whole sample & $23$ & $0.098$ & $24$ & $0.103$\\
\hline
\end{tabular}
\tablefoot{Number of possible isolated stars (with respect to the closest star clusters, OB associations and massive stars) given the threshold distance of $d_{thresh,1}=74$ pc and $d_{thresh,2}=204$ pc. Method $1$ takes into account only the de-projected positions of stars and clusters, while Method $2$ considers clusters whose age difference with the star is $5$~Myr at most. The frequencies are normalized to the total sample of stars.}
\end{table*}

Another criterion for the isolation is that no other massive and young object is found in the vicinity of each target. Following \citet{DorigoJones2020} and \citet{Renzo2019}, we also investigated the isolation of each target with respect to other massive and young stars from our sample. We decided to consider the Method $1$ and the two threshold distances as constraint for the isolation from other massive stars. Table~\ref{tab:Isolation stars} shows the number of objects that have no known star clusters and other massive and young stars nearby for different threshold distances. The new constraint implies a reduction in the isolation frequencies. However, it can be seen that, whatever the method of distances and the threshold value, the final frequency of potentially isolated objects is never zero.

This final sample may contain two types of objects: stars that could have been formed in-situ in the field in isolation (or with a small population of low-mass stars around them) and possible runaways that traveled for a larger period of time than the travel times that we considered or with higher velocities.

\section{Discussion}\label{sec:Discussion}

We investigated the environment around $234$ young massive star candidates in order to study the phenomenon of possible isolated massive star formation. In the previous section we examined the possibility that some of our targets could be runaways from known star clusters or OB associations. Whether runaway objects are produced via binary supernova, dynamical ejection, or subcluster ejection mechanisms, their peculiar velocity in the galaxy plane is at least $24~\text{km}~\text{s}^{-1}$ \citep{DorigoJones2020}. Considering this velocity as typical velocity of runaways, we set $d_{thresh,1}=74$~pc as the first threshold distance a runaway can typically travel, finding with Method $1$ that $68\%\pm6\%~(=159/234)$ of the stars in the sample lie at a larger distance and have no known young star clusters/OB associations around them.

However, there is evidence that some stars could reach very high runaway velocities \citep{DorigoJones2020,Sana2022}. In order to remove possible fast runaways from our sample, we adopted $d_{thresh,2}=204$~pc as our second threshold distance. As a consequence, the fraction of possible isolated stars with respect to star clusters and OB associations using Method $1$ decreases to $34\%\pm4\%~(=79/234)$.

The fact that we set the minimum threshold ($d_{thresh,1}$) to three times the distance a runaway star can travel in $1$~Myr at the minimum $2$D velocity not only enables the selection of stars potentially moving at higher velocities or for longer times, but also accounts for the $\sim0.5$~Mpc uncertainty in the distance of NGC $4242$ \citep{Sabbi2018}. For very high velocities, the number of very fast runaways drops sharply with the velocity \citep{Perets2012, DorigoJones2020}. Increasing $d_{thresh,2}$ would lead to a large number of connections between stars and clusters, not all of which  are necessarily true. In fact, we expect to miss only few faster runaways, given their low number. Moreover, these two threshold distances allowed us not only to identify runaway candidates traveling for only $1$~Myr at the two constant velocities, but also runaways with a combination of velocities and travel times.

The possible connections between stars and stars cluster were based only on the projected distance. If instead one looks also at the similarity in age between stars and star clusters, the number of possible physical connections decreases.
We assumed a maximum age difference between a star and a star cluster of $5$~Myr. This age difference reflects the age spread that \citet{Berlanas2020} and \citet{Schneider2018} found for massive stars in Cygnus OB$2$ and R$136$, respectively. Moreover, $5$~Myr corresponds to the maximum age error we obtained for our sample, and is consistent with the age discrepancy of $0.2$~dex found by \citet{Stevance2020} between single-star isochrone and binary SED fitting with BPASS (Binary Population and Spectral Synthesis). This uncertainty will affect only Method $2$, while the results from Method $1$ will be unaffected.

Ten (for $d_{thresh,1}$) or twenty-four (for $d_{thresh,2}$) stars that we previously associated with star clusters or OB associations as possible runaways, differ more than $5$~Myr from those clustering, making these links less likely to be physical. The resulting fractions of possible isolated objects using Method $2$ become $72\%\pm6\%~(=169/234)$ and $44\%\pm4\%~(=103/234)$ for $d_{thresh,1}$ and $d_{thresh,2}$, respectively.

A further constraint is the isolation with respect to other massive stars outside of clusters. Depending on the threshold distance and method, the frequency can vary from a minimum of $9.8\%\pm2\%~(=23/234$, for $d_{thresh,2}$ and Method $1$) to a maximum of $34.6\%\pm3\%~(=81/234$, for $d_{thresh,1}$ and Method $2$). The errors are for Poisson statistics only.

Observations suggest that $4\%\pm2\%$ of all O-stars in the Milky Way appear to be isolated from known star clusters/OB associations and can not be interpreted as runaways \citep{deWit2005}. This result is consistent with \citet{Vargas-Salazar2020} who found that $4-5\%$ of O-type stars in the field of the Small Magellanic Cloud appear to be isolated from star clusters/OB associations, but these authors also observed small clustering of low-mass stars around these seemingly isolated massive stars. \citet{Parker&Goodwin2007} found that the fraction of sparse star clusters ($M<100M_{\odot})$ with only one massive O-type star can vary between $1.3\%$ and $4.6\%$ for a cluster mass function exponent of $\beta=1.7$ and $\beta=2$, respectively.

In order to compare our findings with those of \citet{deWit2005}, we computed the expected number of massive stars ($M_{ini}\geq15 M_{\odot}$) that have not undergone a SN explosion at $10$~Myr. 
We computed the SFR(UV) directly from the the GALEX FUV image \citep{Lee2009} in the F$275$W HST footprint, using the equation presented in \citet{Kennicutt1998}, a distance of $5.3$~Mpc, and correcting for the Galactic foreground extinction from \citet{Schlafly2011}. These resulted in a SFR(UV)=$0.038 M_{\odot}~\text{yr}^{-1}$. We corrected this star formation rate for the intrinsic mean extinction found in Section~\ref{sec:Age separation}. The final star formation rate is SFR(UV)=$0.057 M_{\odot}~\text{yr}^{-1}$. We also estimated the SFR(H$\alpha$) from the Bok image in the F$275$W HST footprint, following the flux calibration described in \citet{Kennicutt2008} and using the equation presented in \citet{Kennicutt1998}, applying the same distance and foreground and intrinsic extinction corrections (using the mean extinction from Section~\ref{sec:Age separation} using solar metallicity and a MW-like extinction law), and using the star-to-gas extinction conversion from \citet{Calzetti2000}. We found that the SFR(H$\alpha$) is $57\%$ lower than the  SFR(UV), in agreement with the value found by \citet{Lee2009}. Different studies of H\,II regions in galaxies with different metallicities, masses and SFRs have shown that the escape fraction of the ionizing photons can reach values of the order of $67\%$ \citep{DellaBruna2021,Ramambason2022,Egorov2018,Gerasimov2022,Pellegrini2012}. The large escape fraction can lead to an underestimation of the SFR(H$\alpha$) of up to a factor of three. Therefore, we decided to assume a constant SFR=$0.057 M_{\odot}~\text{yr}^{-1}$ from the GALEX FUV over the last $100$~Myr to compute the expected number of massive stars in our HST footprint.
We computed the expected number of massive stars not yet exploded as supernova using the equation presented by \citet{Kim2013}:
\begin{equation}
    \mathcal{P}_{\rm *}=\frac{\dot{M}_*}{m_*}\Delta t
\end{equation}
where $\dot{M}_*=0.057 M_{\odot}\text{yr}^{-1}$ is the dust-corrected SFR(UV), $\Delta t=10$~Myr is our age range, and $m_*=812 M_{\odot}$ is the total mass in massive stars that have not yet exploded as supernova. The latter was computed using a Kroupa IMF \citep{Kroupa2001} and a mass range $15-18.88 M_{\odot}$. These two masses are the lower initial mass for our sample and the upper limit for a star that survives for $10$~Myr without exploding. 
We found that the expected number of massive stars in the field and clusters for our mass and age limits to be $\sim 707$ objects. When we use the expected total number of massive stars, we find that the fraction of possible isolated objects is $3.2 \pm 0.7\% - 11.5 \pm 1.3\%$. The value $11.5\%$ that we find for a threshold of $74$~pc is a factor $3$ larger than what \citet{deWit2005} found using a $65$~pc threshold.

We re-computed the fraction of candidate isolated massive stars using the stellar parameters obtained in Appendix \ref{appendix} for the three different setups described in Section \ref{sec:Age separation}.
For solar metallicity and SMC-like extinction the fractions vary between $2.2 \pm 0.5\% - 7.6 \pm 0.9\%$ due to a larger average extinction and consequently higher dust-corrected SFR(UV), while the fractions become higher if we assume a SMC-like metallicity with MW-like extinction law ($3.3 \pm 0.6\% - 12.0 \pm 1.2\%$). Finally, the combination of a SMC metallicity with the SMC-like extinction law gives fractions in between $2.1 \pm 0.4\% - 7.6 \pm 0.7\%$ (see Appendix \ref{appendix}).

We note that our fractions of isolated massive star candidates are upper limits. In fact, the cluster catalog contains only cluster candidates which were visually inspected and whose effective radius is at least $1$ parsec \citep[see][]{Adamo2017}, missing potentially unresolved clusters with sizes of $1$ parsec and below. The presence of these compact star clusters can reduce the number of possible isolated objects. On the other hand, this is mitigated by the requirement of avoiding other massive stars around each target. Moreover, the distinction between possible runaways from star clusters and isolated objects is based only on position information and not on local velocities for our targets. In fact, the runaway nature of a star can be tested via spectroscopic radial velocities and astrometric tangential velocities in the MW and SMC, but this is not feasible in a galaxy at the distance of NGC $4242$. Therefore, some of our final targets could still be very fast runaways or could have traveled longer from their original birthplace. Finally, some of the isolated candidates may be surrounded by a non-detected stellar clustering of low-mass stars \citep{Vargas-Salazar2020} that we are not able to detect. Conversely, we may be missing truly isolated stars that are seen to lie close to clusters or OB associations in projection.

\section{Conclusions}\label{sec:Conclusions}

In summary, we carried out a galaxy-wide search of massive stars throughout the entire galaxy NGC $4242$ using far and mid- ultraviolet data from HST to find young massive stars that are isolated with respect to star clusters, OB associations and other massive stars. We first presented the identification of possible single young massive field stars exploiting the F$150$LP and F$218$W fluxes from ACS/SBC and WFC3 from HST, identifying $234$ candidates. We then analyzed the surroundings of these targets, looking for the presence of young ($\leq 15$~Myr) star clusters and OB associations identified in the LEGUS survey \citep{Adamo2017}. In fact, massive stars can be ejected from star clusters or OB associations becoming runaways or walkaways.

We set two threshold distances, $74$ and $204$~pc. The former corresponds to three times the distance a massive star can travel in $1$~Myr with the minimum $2$D velocity of $24~\text{km}~\text{s}^{-1}$, while the latter corresponds to the distance a fast runaway can travel in the same time interval with a velocity of $200~\text{km}~\text{s}^{-1}$. Every star that has a star cluster or OB association closer than the threshold distance could be a potential runaway from that star group. We decided to analyze the isolation of our targets based not only on their positions, but considering also the similarity in age between stars and star clusters/OB associations. In this second method, we looked at the spatial distribution of star clusters and OB associations whose age differs less than $5$~Myr from the estimated age of the selected massive star.

The surroundings of the possible isolated objects were then analyzed, looking for the presence of other massive stars. Depending on the method and on the threshold distance, we find a frequency  for isolation in the field of NGC $4242$ that varies from $9.8\%$ to $34.6\%$, using $d_{thresh,2}$ and $d_{thresh,1}$, respectively. These fractions reduce to $3.2 \pm 0.7\% - 11.5 \pm 1.3\%$ when computed with respect to the total expected population of massive stars in the galaxy, with the latter a factor three larger than what \citet{deWit2005} found for the Milky Way. However, our fractions are upper limits to the fraction of truly isolated massive stars. In fact, we cannot probe the possible clustering of low-mass stars around our targets, i.e. we cannot exclude the presence of very sparse star clusters. The fraction of seemingly isolated massive stars around which faint clustering of low-mass stars were found is $4-5\%$ in the Small Magellanic Cloud \citep{Vargas-Salazar2020}. Moreover, our results can still be affected by the presence of very fast runaways or objects that have traveled for a longer period of time. Conversely, we may be missing truly isolated stars located in close apparent proximity to star clusters due to projection effects.

This is the first study that searched for isolated massive stars in a galaxy beyond the immediate neighborhood of the Milky Way. Overall, we find a fraction of potentially isolated massive stars similar to the early work for the Milky Way and Magellanic Clouds. Future studies at higher resolution are needed to elucidate this issue further.   

\begin{acknowledgements}
We thank the anonymous referee for the constructive comments that helped improving the paper.
The HST observations used in this paper are associated with program No. 16316 and No. 13364.
This work is based on observations obtained with the NASA/ESA Hubble Space Telescope, at the Space Telescope Science Institute, which is operated by the Association of Universities for Research in Astronomy, Inc., under NASA contract NAS 5-26555.

E.S. is supported by the international Gemini Observatory, a program of NSF NOIRLab, which is managed by the Association of Universities for Research in Astronomy (AURA) under a cooperative agreement with the U.S. National Science Foundation, on behalf of the Gemini partnership of Argentina, Brazil, Canada, Chile, the Republic of Korea, and the United States of America. J.S.G acknowledges grant support under these GO  programs from the Space Telescope Science Institute which is operated by the Association of Universities for Research in Astronomy, Inc., under NASA contract NAS 5-26555. A.A acknowledges support from Vetenskapsr\aa det 2021-05559. RSK acknowledges financial support from the ERC via Synergy Grant ``ECOGAL'' (project ID 855130),  from the German Excellence Strategy via the Heidelberg Cluster ``STRUCTURES'' (EXC 2181 - 390900948), and from the German Ministry for Economic Affairs and Climate Action in project ``MAINN'' (funding ID 50OO2206).  RSK also thanks the 2024/25 Class of Radcliffe Fellows for highly interesting and stimulating discussions. 
The Legacy Surveys consist of three individual and complementary projects: the Dark Energy Camera Legacy Survey (DECaLS; Proposal ID \#2014B-0404; PIs: David Schlegel and Arjun Dey), the Beijing-Arizona Sky Survey (BASS; NOAO Prop. ID \#2015A-0801; PIs: Zhou Xu and Xiaohui Fan), and the Mayall z-band Legacy Survey (MzLS; Prop. ID \#2016A-0453; PI: Arjun Dey). DECaLS, BASS and MzLS together include data obtained, respectively, at the Blanco telescope, Cerro Tololo Inter-American Observatory, NSF’s NOIRLab; the Bok telescope, Steward Observatory, University of Arizona; and the Mayall telescope, Kitt Peak National Observatory, NOIRLab. Pipeline processing and analyses of the data were supported by NOIRLab and the Lawrence Berkeley National Laboratory (LBNL). The Legacy Surveys project is honored to be permitted to conduct astronomical research on Iolkam Du’ag (Kitt Peak), a mountain with particular significance to the Tohono O’odham Nation.

\end{acknowledgements}
\bibliographystyle{aa}
\bibliography{references}

\begin{appendix}

\section{The effect of different metallicity and dust models on the isolation results}\label{appendix}
Following the mass-metallicity relation from \citet{Galazzi2005} and given the mass $M = 1.2\cdot10^9 ~M_{\odot}$, we decided to explore how our results change when we select stars and fit their SEDs using isochrones computed for $Z=0.004$ (SMC-like metallicity). Using the foreground galactic extinction $A_V=0.033$ mag \citep{Schlafly2011}, the new magnitude cuts for $M = 15 M_{\odot}$ become $25.2$, $21.4$, $22.5$ and $22.1$ mag for F$814$W, F$150$LP, F$275$W and F$218$W filters, respectively. Following the analysis method from Section \ref{sec:Age separation} and the extinction law from \citet{Gordon2016} with $R_V=3.1$ and $f_A=1.0$, we found $306$ possible massive and young stars outside known clusters and associations. The fainter magnitude cuts resulted in a bigger sample of candidates. We re-ran the isochrone-fitting code and found an average age of $8$~Myr and a higher average extinction ($A_V=0.19$ mag) than the results of Section \ref{sec:Age separation}. Tables \ref{Frequency SMC metallicity and MW extinction law}, \ref{Isolation stars SMC metallicity and MW extinction law} show the final results for the isolation with respect to cluster and OB associations and the isolation considering also massive stars. We find that the numbers of isolated candidates (and the relative frequencies) stay close to the results in Section \ref{sec:Discussion}. At this metallicity, stars with $M_{ini}=19.96M_{\odot}$ did not yet explode as SNe, resulting in a total number of expected massive stars of $\sim 876$ for a corrected star formation rate of SFR(UV)=$0.059M_{\odot} ~\text{yr}^{-1}$. Thus, the fraction for the total isolation varies between a minimum of $3.3 \pm 0.6 \%$ to a maximum of $12.0 \pm 1.2\%$, slightly higher than the results from Section \ref{sec:Discussion}.

A second unknown factor that can affect our analysis is the presence or the absence of the $2175$\text{\AA} UV-bump. We present an analysis using solar metallicity and SMC-like extinction law using $R_V=2.74$ \citep{Gordon2003}. The resulting sample of possible massive and young field stars is composed of $235$ objects. The SMC-like extinction law makes the sample slightly younger with a median age of $4.4$~Myr and redder than the MW-like extinction law results. Even if the absence of the $2175$\text{\AA} UV-bump could affect the Method $2$, the isolation analysis shows no/small changes. In fact, the largest variation between the isolation fraction with and without the $2175$\text{\AA} dust-bump is of the order of $1\%$ (see Table \ref{Frequency solar metallicity and SMC extinction law}, \ref{Isolation stars solar metallicity and SMC extinction law} for the isolation fraction using $Z=0.02$ and $R_V=2.74$). However, the higher average extinction ($A_V = 0.19$ mag) results in a higher corrected star formation rate, SFR(UV)=$0.085M_{\odot}~\text{yr}^{-1}$, and a larger number of expected massive stars ($\sim 1052$). The resulting fractions can vary in the smaller range $2.2 \pm 0.5 - 7.6 \pm 0.9\%$.

Finally, it is straightforward to model our isochrone-fitting technique using a SMC metallicity and SMC-like extinction law. We found $306$ possible massive and young field stars: the average age becomes older ($5.5$~Myr) and the mean extinction $A_V=0.22$ mag. Even if the new set of analyses describes a sample older than the original one, we found only very small changes in the fraction of isolated stars (see Tables \ref{Frequency SMC metallicity and SMC extinction law}, \ref{Isolation stars SMC metallicity and SMC extinction law}). The higher maximum initial mass to not undergo a SN explosion at $10$~Myr at SMC metallicity and the higher extinction result in a corrected star formation rate of $0.094M_{\odot}~\text{yr}^{-1}$ and an expected number of massive stars of $\sim 1382$. The resulting fraction can vary between $2.1 \pm 0.4 - 7.6 \pm 0.7\%$.

\onecolumn
\begin{table*}
\centering
\small
\setlength{\tabcolsep}{3pt}
\renewcommand{\arraystretch}{0.9}
\caption{Number of possible isolated stars relative to nearby clusters or OB associations (SMC-like metallicity and MW-like extinction law setup).}
\label{Frequency SMC metallicity and MW extinction law}
\centering
\begin{tabular}{ccccc}
\hline \hline
Sample & \multicolumn{2}{c}{Method $1$} & \multicolumn{2}{c}{Method $2$}\\
\hline \hline
$d_{thresh,1}=74$pc & Number & Frequency & Number & Frequency\\
\hline
only clusters\\
dof=$4$ & $93$ & $0.304$ & $104$ & $0.340$ \\
dof=$3$ & $117$ & $0.382$ & $123$ & $0.402$\\
dof=$2$ & $23$ & $0.075$ & $23$ & $0.075$\\
whole sample & $233$ & $0.761$ & $250$ & $0.817$\\
\hline
only OB associations \\
dof=$4$ & $111$ & $0.363$ & $121$ & $0.395$\\
dof=$3$ & $128$ & $0.418$ & $133$ & $0.435$\\
dof=$2$ & $18$ & $0.059$ & $21$ & $0.069$\\
whole sample & $257$ & $0.840$ & $275$ & $0.899$\\
\hline
clusters + OB associations \\
dof=$4$ & $89$ & $0.291$ & $102$ & $0.333$\\
dof=$3$ & $112$ & $0.366$ & $119$ & $0.389$\\
dof=$2$ & $16$ & $0.052$ & $19$ & $0.062$\\
whole sample & $217$ & $0.709$ & $240$ & $0.784$\\
\hline \hline
$d_{thresh,2}=204$pc & Number & Frequency & Number & Frequency\\
\hline
only clusters \\
dof=$4$ & $44$ & $0.144$ & $59$ & $0.193$\\
dof=$3$ & $74$ & $0.242$ & $87$ & $0.284$\\
dof=$2$ & $12$ & $0.039$ & $15$ & $0.049$\\
whole sample & $130$ & $0.425$ & $161$ & $0.526$\\
\hline
only OB associations \\
dof=$4$ & $89$ & $0.291$ & $105$ & $0.343$\\
dof=$3$ & $117$ & $0.382$ & $127$ & $0.415$\\
dof=$2$ & $13$ & $0.042$ & $18$ & $0.059$\\
whole sample & $219$ & $0.716$ & $250$ & $0.817$\\
\hline
clusters + OB associations \\
dof=$4$ & $34$ & $0.111$ & $52$ & $0.170$\\
dof=$3$ & $65$ & $0.212$ & $79$ & $0.258$\\
dof=$2$ & $6$ & $0.020$ & $13$ & $0.042$\\
whole sample & $105$ & $0.343$ & $144$ & $0.471$\\
\hline
\end{tabular}
\tablefoot{Number and frequency of possible isolated stars (with respect to the closest star clusters or OB associations) given the threshold distance of $d_{thresh,1}=74$ pc and $d_{thresh,2}=204$ pc. Method $1$ takes into account only the de-projected positions of stars and clusters, while Method $2$ considers clusters whose age difference is $5$ Myr at most. The number of stars is $306$ using a SMC-like metallicity and a MW-like extinction law.}

\vspace{2mm}

\setlength{\tabcolsep}{3pt}
\renewcommand{\arraystretch}{0.9}
\centering
\small
\caption{Number of possible isolated stars relative to nearby clusters, OB associations, and massive stars (SMC-like metallicity and MW-like extinction law setup).}
\label{Isolation stars SMC metallicity and MW extinction law}
\centering
\begin{tabular}{ccccc}
\hline \hline
Sample & \multicolumn{2}{c}{Method $1$} & \multicolumn{2}{c}{Method $2$} \\
\hline
\ & Number & Frequency & Number & Frequency\\
\hline \hline
clusters + OB associations ($d_{thresh,1}$) \\
and stars ($d_{thresh,1}$) isolation\\
dof=$4$ & $39$ & $0.127$ & $39$ & $0.127$\\
dof=$3$ & $58$ & $0.190$ & $59$ & $0.193$\\
dof=$2$ & $7$ & $0.023$ & $7$ & $0.023$\\
whole sample & $104$ & $0.340$ & $105$ & $0.343$\\
\hline
clusters + OB associations ($d_{thresh,1}$) \\
and stars ($d_{thresh,2}$) isolation\\
dof=$4$ & $12$ & $0.039$ & $12$ & $0.039$\\
dof=$3$ & $25$ & $0.082$ & $25$ & $0.082$\\
dof=$2$ & $3$ & $0.010$ & $3$ & $0.010$\\
whole sample & $40$ & $0.131$ & $40$ & $0.131$\\
\hline
clusters + OB associations ($d_{thresh,2}$) \\
and stars ($d_{thresh,1}$) isolation\\
dof=$4$ & $20$ & $0.065$ & $24$ & $0.078$\\
dof=$3$ & $36$ & $0.118$ & $41$ & $0.134$\\
dof=$2$ & $3$ & $0.010$ & $5$ & $0.0163$\\
whole sample & $59$ & $0.193$ & $70$ & $0.229$\\
\hline
clusters + OB associations ($d_{thresh,2}$) \\
and stars ($d_{thresh,2}$) isolation\\
dof=$4$ & $9$ & $0.029$ & $9$ & $0.029$\\
dof=$3$ & $17$ & $0.056$ & $18$ & $0.059$\\
dof=$2$ & $3$ & $0.010$ & $3$ & $0.010$\\
whole sample & $29$ & $0.095$ & $30$ & $0.098$\\
\hline
\end{tabular}
\tablefoot{Number of possible isolated stars (with respect to the closest star clusters, OB associations and massive stars) given the threshold distance of $d_{thresh,1}=74$ pc and $d_{thresh,2}=204$ pc. Method $1$ takes into account only the de-projected positions of stars and clusters, while Method $2$ considers clusters whose age difference is $5$ Myr at most. The number of stars is $306$ using a SMC-like metallicity and a MW-like extinction law.}
\end{table*}
\twocolumn

\onecolumn
\begin{table*}
\centering
\small
\setlength{\tabcolsep}{3pt}
\renewcommand{\arraystretch}{0.9}
\caption{Number of possible isolated stars relative to nearby clusters or OB associations (solar metallicity and SMC-like extinction law setup).}
\label{Frequency solar metallicity and SMC extinction law}
\centering
\begin{tabular}{ccccc}
\hline \hline
Sample & \multicolumn{2}{c}{Method $1$} & \multicolumn{2}{c}{Method $2$} \\
\hline \hline
$d_{thresh,1}=74$pc & Number & Frequency & Number & Frequency\\
\hline
only clusters\\
dof=$4$ & $70$ & $0.298$ & $70$ & $0.298$ \\
dof=$3$ & $87$ & $0.370$ & $91$ & $0.387$\\
dof=$2$ & $19$ & $0.081$ & $20$ & $0.085$\\
whole sample & $176$ & $0.749$ & $181$ & $0.770$\\
\hline
only OB associations \\
dof=$4$ & $81$ & $0.345$ & $83$ & $0.353$\\
dof=$3$ & $93$ & $0.396$ & $95$ & $0.404$\\
dof=$2$ & $15$ & $0.064$ & $16$ & $0.068$\\
whole sample & $189$ & $0.804$ & $194$ & $0.826$\\
\hline
clusters + OB associations \\
dof=$4$ & $66$ & $0.281$ & $66$ & $0.281$\\
dof=$3$ & $81$ & $0.345$ & $86$ & $0.366$\\
dof=$2$ & $13$ & $0.055$ & $15$ & $0.064$\\
whole sample & $160$ & $0.681$ & $167$ & $0.711$\\
\hline \hline
$d_{thresh,2}=204$pc & Number & Frequency & Number & Frequency\\
\hline
only clusters \\
dof=$4$ & $35$ & $0.149$ & $41$ & $0.174$\\
dof=$3$ & $58$ & $0.247$ & $70$ & $0.298$\\
dof=$2$ & $9$ & $0.038$ & $11$ & $0.047$\\
whole sample & $102$ & $0.434$ & $122$ & $0.519$\\
\hline
only OB associations \\
dof=$4$ & $64$ & $0.272$ & $68$ & $0.289$\\
dof=$3$ & $85$ & $0.362$ & $91$ & $0.387$\\
dof=$2$ & $11$ & $0.047$ & $13$ & $0.055$\\
whole sample & $160$ & $0.681$ & $172$ & $0.732$\\
\hline
clusters + OB associations \\
dof=$4$ & $26$ & $0.111$ & $33$ & $0.140$\\
dof=$3$ & $50$ & $0.213$ & $64$ & $0.272$\\
dof=$2$ & $4$ & $0.17$ & $8$ & $0.034$\\
whole sample & $80$ & $0.340$ & $105$ & $0.447$\\
\hline
\end{tabular}
\tablefoot{Number and frequency of possible isolated stars (with respect to the closest star clusters or OB associations) given the threshold distance of $d_{thresh,1}=74$ pc and $d_{thresh,2}=204$ pc. Method $1$ takes into account only the de-projected positions of stars and clusters, while Method $2$ considers clusters whose age difference is $5$ Myr at most. The number of stars is $235$ using a solar metallicity and a SMC-like extinction law.}

\vspace{2mm}

\centering
\small
\caption{Number of possible isolated stars relative to nearby clusters, OB associations, and massive stars (solar metallicity and SMC-like extinction law setup).}
\label{Isolation stars solar metallicity and SMC extinction law}
\centering
\begin{tabular}{ccccc}
\hline \hline
Sample & \multicolumn{2}{c}{Method $1$} & \multicolumn{2}{c}{Method $2$}\\
\hline
\ & Number & Frequency & Number & Frequency\\
\hline \hline
clusters + OB associations ($d_{thresh,1}$) \\
and stars ($d_{thresh,1}$) isolation\\
dof=$4$ & $31$ & $0.132$ & $31$ & $0.132$\\
dof=$3$ & $41$ & $0.174$ & $42$ & $0.179$\\
dof=$2$ & $6$ & $0.026$ & $7$ & $0.0298$\\
whole sample & $78$ & $0.332$ & $80$ & $0.340$\\
\hline
clusters + OB associations ($d_{thresh,1}$) \\
and stars ($d_{thresh,2}$) isolation\\
dof=$4$ & $6$ & $0.026$ & $6$ & $0.026$\\
dof=$3$ & $20$ & $0.085$ & $20$ & $0.085$\\
dof=$2$ & $2$ & $0.009$ & $2$ & $0.009$\\
whole sample & $28$ & $0.119$ & $28$ & $0.119$\\
\hline
clusters + OB associations ($d_{thresh,2}$) \\
and stars ($d_{thresh,1}$) isolation\\
dof=$4$ & $15$ & $0.064$ & $17$ & $0.072$\\
dof=$3$ & $29$ & $0.123$ & $32$ & $0.136$\\
dof=$2$ & $2$ & $0.009$ & $5$ & $0.021$\\
whole sample & $46$ & $0.196$ & $54$ & $0.230$\\
\hline
clusters + OB associations ($d_{thresh,2}$) \\
and stars ($d_{thresh,2}$) isolation\\
dof=$4$ & $4$ & $0.017$ & $4$ & $0.017$\\
dof=$3$ & $17$ & $0.072$ & $17$ & $0.072$\\
dof=$2$ & $2$ & $0.009$ & $2$ & $0.009$\\
whole sample & $23$ & $0.098$ & $23$ & $0.098$\\
\hline
\end{tabular}
\tablefoot{Number of possible isolated stars (with respect to the closest star clusters, OB associations and massive stars) given the threshold distance of $d_{thresh,1}=74$ pc and $d_{thresh,2}=204$ pc. Method $1$ takes into account only the de-projected positions of stars and clusters, while Method $2$ considers clusters whose age difference is $5$ Myr at most. The number of stars is $235$ using a solar metallicity and a SMC-like extinction law.}
\end{table*}
\twocolumn

\onecolumn
\begin{table*}
\centering
\small
\setlength{\tabcolsep}{3pt}
\renewcommand{\arraystretch}{0.9}
\caption{Number of possible isolated stars relative to nearby clusters or OB associations (SMC-like metallicity and SMC-like extinction law setup).}
\label{Frequency SMC metallicity and SMC extinction law}
\centering
\begin{tabular}{ccccc}
\hline \hline
Sample & \multicolumn{2}{c}{Method $1$} & \multicolumn{2}{c}{Method $2$} \\
\hline \hline
$d_{thresh,1}=74$pc & Number & Frequency & Number & Frequency\\
\hline
only clusters\\
dof=$4$ & $93$ & $0.304$ & $98$ & $0.320$ \\
dof=$3$ & $117$ & $0.382$ & $126$ & $0.412$\\
dof=$2$ & $23$ & $0.075$ & $23$ & $0.075$\\
whole sample & $233$ & $0.761$ & $247$ & $0.807$\\
\hline
only OB associations \\
dof=$4$ & $111$ & $0.363$ & $117$ & $0.382$\\
dof=$3$ & $128$ & $0.418$ & $133$ & $0.435$\\
dof=$2$ & $18$ & $0.059$ & $20$ & $0.065$\\
whole sample & $257$ & $0.840$ & $270$ & $0.882$\\
\hline
clusters + OB associations \\
dof=$4$ & $89$ & $0.291$ & $94$ & $0.307$\\
dof=$3$ & $112$ & $0.366$ & $122$ & $0.399$\\
dof=$2$ & $16$ & $0.052$ & $18$ & $0.059$\\
whole sample & $217$ & $0.709$ & $234$ & $0.765$\\
\hline \hline
$d_{thresh,2}=204$pc & Number & Frequency & Number & Frequency\\
\hline
only clusters \\
dof=$4$ & $44$ & $0.144$ & $52$ & $0.170$\\
dof=$3$ & $74$ & $0.242$ & $93$ & $0.304$\\
dof=$2$ & $12$ & $0.039$ & $14$ & $0.046$\\
whole sample & $130$ & $0.425$ & $159$ & $0.520$\\
\hline
only OB associations \\
dof=$4$ & $89$ & $0.291$ & $100$ & $0.327$\\
dof=$3$ & $117$ & $0.382$ & $127$ & $0.415$\\
dof=$2$ & $13$ & $0.042$ & $16$ & $0.052$\\
whole sample & $219$ & $0.716$ & $243$ & $0.794$\\
\hline
clusters + OB associations \\
dof=$4$ & $34$ & $0.111$ & $43$ & $0.141$\\
dof=$3$ & $65$ & $0.212$ & $85$ & $0.278$\\
dof=$2$ & $6$ & $0.020$ & $10$ & $0.033$\\
whole sample & $105$ & $0.343$ & $138$ & $0.451$\\
\hline
\end{tabular}
\tablefoot{Number and frequency of possible isolated stars (with respect to the closest star clusters or OB associations) given the threshold distance of $d_{thresh,1}=74$ pc and $d_{thresh,2}=204$ pc. Method $1$ takes into account only the de-projected positions of stars and clusters, while Method $2$ considers clusters whose age difference is $5$ Myr at most. The number of stars is $306$ using a SMC-like metallicity and a SMC-like extinction law.}

\vspace{2mm}

\centering
\small
\caption{Number of possible isolated stars relative to nearby clusters, OB associations, and massive stars (SMC-like metallicity and SMC-like extinction law setup).}
\label{Isolation stars SMC metallicity and SMC extinction law}
\centering
\begin{tabular}{ccccc}
\hline \hline
Sample & \multicolumn{2}{c}{Method $1$} & \multicolumn{2}{c}{Method $2$} \\
\hline
\ & Number & Frequency & Number & Frequency\\
\hline \hline
clusters + OB associations ($d_{thresh,1}$) \\
and stars ($d_{thresh,1}$) isolation\\
dof=$4$ & $39$ & $0.127$ & $39$ & $0.127$\\
dof=$3$ & $58$ & $0.190$ & $59$ & $0.193$\\
dof=$2$ & $7$ & $0.023$ & $7$ & $0.023$\\
whole sample & $104$ & $0.340$ & $105$ & $0.343$\\
\hline
clusters + OB associations ($d_{thresh,1}$) \\
and stars ($d_{thresh,2}$) isolation\\
dof=$4$ & $12$ & $0.039$ & $12$ & $0.039$\\
dof=$3$ & $25$ & $0.082$ & $25$ & $0.082$\\
dof=$2$ & $3$ & $0.010$ & $3$ & $0.010$\\
whole sample & $40$ & $0.131$ & $40$ & $0.131$\\
\hline
clusters + OB associations ($d_{thresh,2}$) \\
and stars ($d_{thresh,1}$) isolation\\
dof=$4$ & $20$ & $0.065$ & $23$ & $0.075$\\
dof=$3$ & $36$ & $0.118$ & $42$ & $0.137$\\
dof=$3$ & $3$ & $0.010$ & $3$ & $0.010$\\
whole sample & $59$ & $0.193$ & $68$ & $0.222$\\
\hline
clusters + OB associations ($d_{thresh,2}$) \\
and stars ($d_{thresh,2}$) isolation\\
dof=$4$ & $9$ & $0.029$ & $9$ & $0.029$\\
dof=$3$ & $17$ & $0.056$ & $18$ & $0.059$\\
dof=$2$ & $3$ & $0.010$ & $3$ & $0.010$\\
whole sample & $29$ & $0.095$ & $30$ & $0.098$\\
\hline
\end{tabular}
\tablefoot{Number of possible isolated stars (with respect to the closest star clusters, OB associations and massive stars) given the threshold distance of $d_{thresh,1}=74$ pc and $d_{thresh,2}=204$ pc. Method $1$ takes into account only the de-projected positions of stars and clusters, while Method $2$ considers clusters whose age difference is $5$ Myr at most. The number of stars is $306$ using a SMC-like metallicity and a SMC-like extinction law.}
\end{table*}
\twocolumn
\end{appendix}

\end{document}